\documentclass[preprint,12pt]{elsarticle}
\usepackage[margin=2.5cm]{geometry}
\usepackage{graphicx}
\usepackage{amssymb}
\usepackage{tabularx}
\usepackage{newtxmath}
\usepackage{mathbbol}
\usepackage{lineno}
\usepackage{hyperref}
\usepackage{soul}

\usepackage{subfiles}
\usepackage{tikz}
\usepackage{pgfplots}
\usepgfplotslibrary{colorbrewer}
\pgfplotsset{compat=newest}
\usetikzlibrary{positioning, calc, backgrounds, shapes.geometric}
\usetikzlibrary{external}
\usepgfplotslibrary{colormaps}

\newcommand{\beq}{\begin{equation}}
\newcommand{\eeq}{\end{equation}}

\usepackage{comment}
\usepackage{amsmath}
\usepackage{lipsum}
\usepackage{graphics}
\usepackage{ulem}

\journal{Arxiv}
\begin{document}

\begin{frontmatter}
\title{Benchmarks for physics-informed data-driven hyperelasticity }

\author{Vahidullah Tac$^{1}$, Kevin Linka$^{2}$, Francisco Sahli-Costabal$^{3}$, Ellen Kuhl$^{2}$ and Adrian Buganza Tepole$^{1,4}$ }

\address{$^1$School of Mechanical Engineering, Purdue University, West Lafayette, USA\\ $^2$Department of Mechanical Engineering, Stanford University,  Stanford, USA\\$^3$Department of Mechanical Engineering, Pontifica Universidad Catolica de Chile, Santiago, Chile\\$^4$Weldon School of Biomedical Engineering, Purdue University, West Lafayette, USA}

\begin{abstract}
Data-driven methods have changed the way we understand and model materials. However, while providing unmatched flexibility, these methods have limitations such as reduced capacity to extrapolate, overfitting, and violation of physics constraints. Recent developments have led to modeling frameworks that automatically satisfy these requirements. Here we review, extend, and compare three promising data-driven methods: Constitutive Artificial Neural Networks (CANN), Input Convex Neural Networks (ICNN), and Neural Ordinary Differential Equations (NODE). Our formulation expands the strain energy potentials in terms of sums of convex non-decreasing functions of invariants and linear combinations of these. The expansion of the energy is shared across all three methods and guarantees the automatic satisfaction of objectivity and polyconvexity, essential within the context of hyperelasticity. To benchmark the methods, we train them against rubber and skin stress-strain data. All three approaches capture the data almost perfectly, without overfitting, and have some capacity to extrapolate. Interestingly, the methods find different energy functions even though the prediction on the stress data is nearly identical. The most notable differences are observed in the second derivatives, which could impact performance of numerical solvers. On the rich set of data used in these benchmarks, the models show the anticipated trade-off between number of parameters and accuracy. Overall, CANN, ICNN and NODE retain the flexibility and accuracy of other data-driven methods without compromising on the physics. These methods are thus ideal options to model arbitrary hyperelastic material behavior.
\end{abstract}


\end{frontmatter}

\section*{Introduction}\label{motiv}

The frontier of biomedical engineering applications such as personalized surgery requires accurate mathematical models of material-specific behavior \cite{lee2020}. Similarly, human-engineered systems based on soft materials also necessitate predictive simulations with high precision \cite{duriez2017soft}.
The materials for these applications are extremely nonlinear and undergo large deformations, e.g. rubber and skin. Yet, despite decades of effort developing constitutive equations for these materials, there still isn't a definitive model for them due to inherent limitations of expert-constructed models \cite{limbert2019skin}. Traditional material modeling restricts the prediction of the mechanical response to a narrow set of functional terms, making it nearly impossible to fully capture the data. On the other hand, data-driven methods such as neural networks are universal approximators that can fit mechanical behavior data of complex response almost perfectly \cite{leshno1993multilayer,peng2020multiscale}. Data-driven methods, on the other hand, have their own drawbacks. Most importantly, considering the response of rubbers and many biological materials as hyperelastic, the mechanical response is fully specified by a scalar potential that has to satisfy the conditions of objectivity and polyconvexity \cite{marsden1994mathematical}. These physics constraints are crucial for enabling robust large-scale simulations, extrapolate from data, and avoid over-fitting \cite{tac2022data}. Yet, off-the-shelf machine learning tools cannot \textit{a priori} satisfy these conditions. We review and refine three very recent developments in physics-informed machine learning that aim at embedding the objectivity and polyconvexity constraints as part of the formulation such that they can be satisfied \textit{a priori}. The methods we consider are three: Constitutive Artificial Neural Netwokrs (CANN) \cite{linka2023new}, Neural Ordinary Differential Equations (NODE) \cite{TAC2022115248}, and Input Convex Neural Networks (ICNN) \cite{klein2022polyconvex,amos2017input}.

Characterizing nonlinear materials like rubber and skin involves testing them under a wide set of deformation modes such as uniaxial, biaxial, shear, and sometimes triaxial deformation. The resulting data collected from these tests are stress-strain curves. The direct approach to leverage machine learning on these data is to directly map between strains and stresses \cite{ghaboussi1998_2}. One problem with this strategy is that objectivity is not preserved. One way of fixing this issue has been to augment the data with superimposed arbitrary rotations \cite{heider2020so, klein2022polyconvex}. This solution does not fully guarantee objectivity, although, as the space of rotations is sampled more and more thoroughly the constraint is more closely satisfied. For closed-form constitutive models, fulfilling objectivity has not been a major hurdle. Experts develop strain energies in terms of invariants of the deformation to satisfy objectivity by default \cite{holzapfel2000,ehret2007polyconvex}. Machine learning methods along these lines have also been proposed \cite{garikipati2020multiresolution,vlassis2020geometric,liu2020,fuhg2022learning}. The challenge of imposing polyconvexity in data-driven methods is more difficult to address. Polyconvexity of the strain energy (with additional growth conditions) is a sufficient condition for the existence of solutions for boundary value problems in hyperelasticity \cite{ball1976convexity}. Polyconvexity is a flexible framework, compatible with unstable material behavior like buckling. This is in contrast with more restrictive notions like convexity of the strain energy with respect to the deformation gradient, which can violate objectivity \cite{schroder2010anisotropic}. Laxer conditions such as rank-one convexity are weaker than polyconvexity and not sufficient for the existence of global minimizers of the strain energy \cite{schroder2003invariant}. Many expert models are based on the notion of polyconvexity (but there are also many examples of popular models which might violate this condition \cite{gao2017convexity}). Different notions of convexity have been considered within data-driven frameworks, but the majority have opted for adding the constraint as a penalty through the loss function \cite{vlassis2020geometric}. 

CANNs are a new method for automated model discovery that borrow their architecture from traditional feed-forward neural networks but use activation functions that preserve convexity. They also prune the connections between the inputs and subsequent nodes such that the final result is a polyconvex strain energy \cite{linka2023new}. ICNNs also build convex functions of invariant inputs using specific activation functions and non-negative weights \cite{as2022mechanics, CHEN2022103993}. NODEs tackle the problem differently by leveraging the monotonicity of trajectories of ordinary differential equations (ODEs) in a single variable to interpolate monotic functions associated with derivatives of a strain energy rather than the energy directly \cite{TAC2022115248}.  Unlike other approaches, CANNs, ICNNs, and NODEs have physics at the core of the formulation to generate constitutive models that satisfy objectivity and polyconvexity \textit{a priori}. Yet, there is a gap in our understanding of how these different methods perform on benchmark datasets, and a general need to benchmark machine learning methods in computational mechanics \cite{lejeune2020mechanical,kobeissi2022enhancing}. 

Rubber modeling was the center of attention for large deformation hyperelasticity in the past century, with tens of constitutive models proposed \cite{dal2021performance}. Recently, advances in soft robotics  has renewed the interest in developing improved high-fidelity simulations of soft robots made of rubbers and other elastomers. For example, applications that aim at produce complex motion such as tentacle grippers, walking soft robots, and rehabilitation soft exoskeletons \cite{rus2015design}, all require precise modeling of the material response. 

Soft tissues made of collagen have remarkable mechanical properties. They show exponential-like stress-strain response and anisotropy. These nonlinearities allow tissues like skin to protect us against environmental harm while allowing interaction and movement \cite{jor2013computational}. The development of constitutive models for soft tissues, and skin in particular, dates to the seminal work by Lanir and Fung \cite{lanir1974two,lanir1983constitutive}, and has resulted in a long list of strain energy functions proposed over the past five decades \cite{limbert2019skin}.  New models are being proposed even today \cite{toaquiza2022anisotropic,chen2020microstructurally}. Despite the rich literature on skin and soft tissue modeling, the complexity of the material response in these materials has prevented the emergence of a categorically superior constitutive model.

The manuscript is organized as follows. In the Methods section we first review the basic equations that describe the mechanical behavior of hyperelastic materials with emphasis on strain energy function expansions that satisfy objectivity and polyconvexity requirements. Then, we show how  CANN, ICNN and NODE architectures can be used to create material models within the considered families of elastic potentials. After training the three methods to datasets on rubber and skin, the Results section explores in detail the ability of the models to interpolate and extrapolate, their robustness with respect to model initialization, the regularity of second derivatives of the energy, and the trade-off between number of parameters and model accuracy. We finally discuss the results in the context of other data-driven efforts for computational mechanics.

\begin{figure*}[h!]
\centering
\includegraphics[width=17cm]{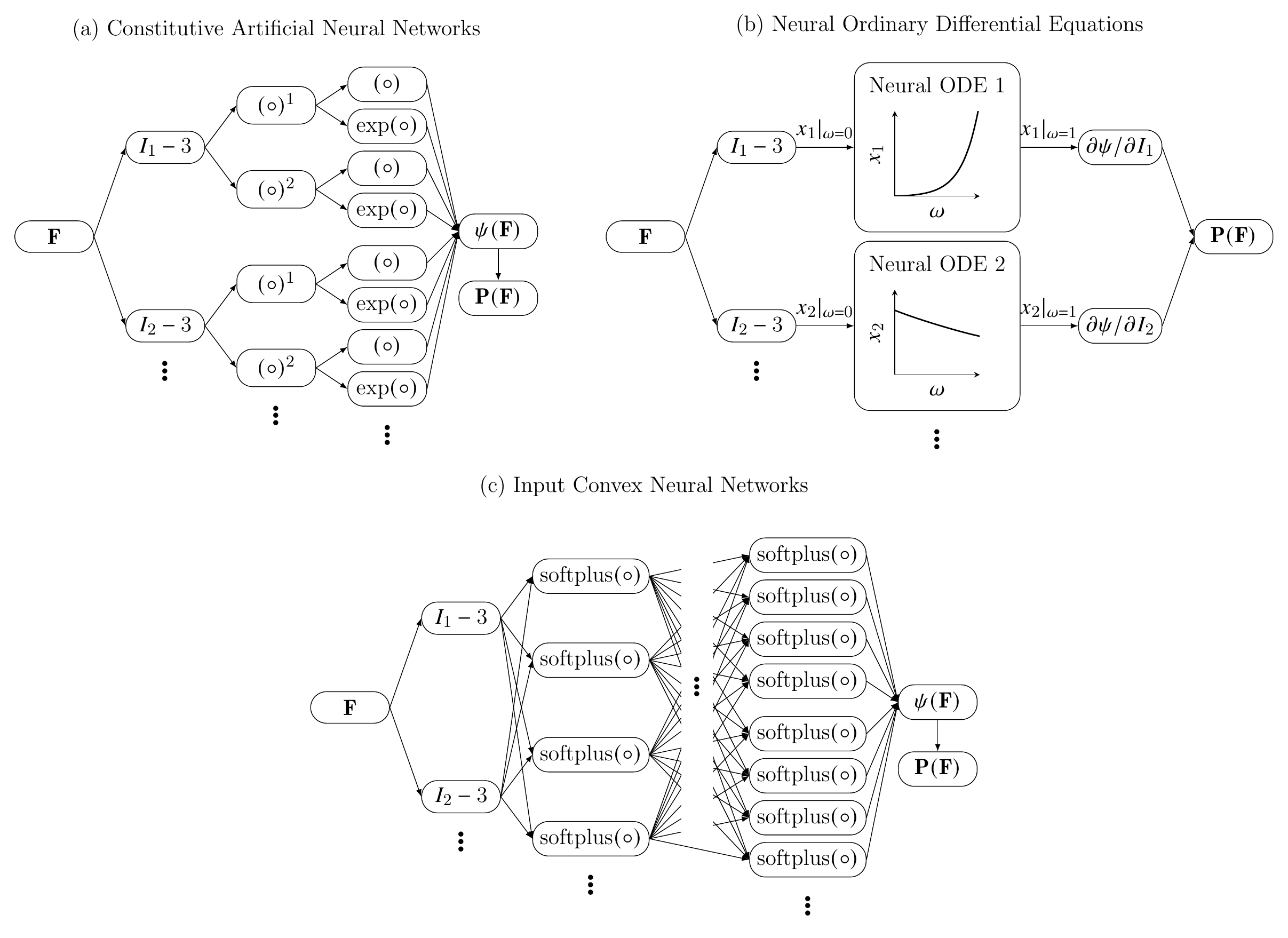} \vspace{-4mm}
\caption{Diagram depicting the training and inference processes of the deep neural network material model.}
\label{fig01} 
\end{figure*}

\section*{Methods}

\subsection*{Polyconvex strain energy density functions}

Consider a motion $\mathbf{\varphi}$, the gradient $\mathbf{F}=\nabla \mathbf{\varphi}$ contains all the local information about the deformation. Within the framework of hyperelasticity, the strain energy function $\psi(\mathbf{F})$ fully defines the material response. Polyconvexity implies that the energy $\psi(\mathbf{F})$ can be expressed as a convex function in the extended domain $\hat{\psi}(\mathbf{F},\mathrm{cof}\,\mathbf{F},\det \mathbf{F})$. Intuitively, this extended domain covers different modes of deformation: $\mathbf{F}$ measures changes in length, $\mathrm{cof}\,\mathbf{F}$ changes in area, and $J=\det \mathbf{F}$  changes in volume. It is difficult to work directly with the deformation gradient and its cofactors as inputs to the strain energy. Instead, the right Cauchy-Green deformation tensor $\mathbf{C}=\mathbf{F}^\top \mathbf{F}$ is used because it does not contain information about superimposed rigid body rotations. Furthermore, objectivity is enforced by working with the invariants 

\begin{equation}
\begin{aligned}
&I_1 = \mathrm{tr}\,\mathbf{C} = \mathbf{C}:\mathbf{I}\\
&I_2 = \frac{1}{2}\left( (\mathrm{tr}\,\mathbf{C})^2 - \mathrm{tr}\,\mathbf{C}^2\right)\\
&I_3 = J^2 = \det \mathbf{C} \\
&I_{4a} = \mathbf{C}:\mathbf{a}_0 \otimes \mathbf{a}_0 \, , \; I_{4s} =\mathbf{C}:\mathbf{s}_0 \otimes \mathbf{s}_0 \, .
\end{aligned}
\label{eq:invariants}
\end{equation}

The last two invariants in eq.(\ref{eq:invariants}) are only relevant for transversely anisotropic materials and depend on the deformation of two  material unit vectors $\mathbf{a}_0$, $\mathbf{s}_0$. For soft tissues, the vectors $\mathbf{a}_0$, $\mathbf{s}_0$ represent collagen fiber bundle orientations. Furthermore, for nearly incompressible materials such as rubbers and skin, the split between volumetric and isochoric parts is often used. The isochoric part of the deformation is $\bar{\mathbf{F}} = J^{-1/3}\mathbf{F}$, with the corresponding deformation tensor  $\bar{\mathbf{C}}=\bar{\mathbf{F}}^\top \bar{\mathbf{F}}$. The isochoric invariants follow 

\begin{equation}
\begin{aligned}
&\bar{I}_1 = J^{-2/3}I_1\\
&\bar{I}_2 = J^{-4/3}I_2\\
&\bar{I}_{4a} = J^{-2/3}I_{4a}\, , \; \bar{I}_{4s} = J^{-2/3}I_{4s}\, .
\end{aligned}
\label{eq:iso_invariants}
\end{equation}

Based on the split between the isochoric and volumetric parts of the deformation, the energy can be additively decomposed into

\begin{equation}
\psi = \psi_{\mathrm{iso}}(\bar{I}_1,\bar{I}_2,\bar{I}_{4a},\bar{I}_{4s})+\psi_{\mathrm{vol}}(J)\, .
\label{eq:psi_iso_vol}
\end{equation}

For polyconvexity to be satisfied in this additive split, one requirement is convexity of $\psi_{\mathrm{vol}}$ and growth conditions $\psi_{\mathrm{vol}} \to \infty$ as $J \to 0$ or $J \to \infty$. In the case of fully incompressible materials, the volumetric part of the strain energy is replaced by $p(J-1)$, where $p$ is a Lagrange multiplier field that enforces $J=1$. In simple loading cases such as uniaxial or biaxial deformation, $p$ can be directly determined from boundary conditions. In addition, for incompressible behavior the isochoric part of the energy becomes a function of the original invariants defined in eq. (\ref{eq:invariants}).

To ensure polyconvexity of $\psi_{\mathrm{iso}}$, recall that this condition implies a function $\hat{\psi}$ convex on $(\mathbf{F},\mathrm{cof}\,\mathbf{F},\det \mathbf{F})$. The invariant $I_1$ is convex in $\mathbf{F}$, while $I_2$ is convex in $\mathrm{cof}\,\mathbf{F}$. The anisotropic invariants $I_{4a}$, $I_{4s}$ are also convex on $\mathbf{F}$. Moreover, the isochoric split preserves polyconvexity of $\bar{I}_1$, $\bar{I}_{4a}$, $\bar{I}_{4s}$,  and a simple scaling with a power of $J$ is enough to maintain polyconvexity of $\bar{I}_2$. Thus, a sufficient large family of polyconvex functions has the form 

\begin{equation}
\psi = \psi_{1}(\bar{I}_1)+\psi_{2}(\bar{I}_2)+\psi_{4a}(\bar{I}_{4a})+\psi_{4s}(\bar{I}_{4s})+\psi_{vol}(J)\, , 
\label{eq:psi_polyconvex_additive1}
\end{equation}

with each of the $\psi_i$ a convex non decreasing function of its argument, while, as mentioned previously, $\psi_{\mathrm{vol}}$ has to be convex and grow to infinity appropriately with changes in $J$. Again, for incompressibility, the last term in (\ref{eq:psi_polyconvex_additive1}) is replaced with the Lagrange multiplier constraint, and the $\psi_i$ terms can be considered as functions of the invariants in eq. (\ref{eq:invariants}).

\subsection*{Stress predictions from a strain energy potential}

Given a strain energy function, the second Piola-Kirchhoff stress follows from the standard Coleman-Noll procedure \cite{marsden1994mathematical},

\begin{equation}
\mathbf{S} = 2 \frac{\partial{\psi}}{{\partial \mathbf{C}}} \, .
\label{eq:S_def}
\end{equation}

Other stress tensors can be easily obtained with push-forward operations, for instance the nominal or first Piola-Kirchhoff stress $\mathbf{P}=\mathbf{F}\mathbf{S}$, or the Cauchy stress $\mathbf{\sigma}=J^{-1}\mathbf{F}\mathbf{S}\mathbf{F}^\top$, which appear in the strong form of linear momentum balance in the reference or deformed configurations respectively. Since the energy is in terms of the invariants, computing the stress requires the standard derivatives 

\begin{equation}
\begin{aligned}
&\frac{\partial I_1}{\partial \mathbf{C}} = \mathbf{I}\\
&\frac{\partial I_2}{\partial \mathbf{C}} = \frac{1}{2}(I_1 \mathbf{I}-\mathbf{C})\\
&\frac{\partial I_3}{\partial \mathbf{C}} = I_3 \mathbf{C}^{-1}\\
&\frac{\partial I_{4a}}{\partial \mathbf{C}} = \mathbf{a}_0\otimes \mathbf{a_0}\, , \;
\frac{\partial I_{4s}}{\partial \mathbf{C}} = \mathbf{s}_0\otimes \mathbf{s_0} \, .
\end{aligned}
\label{eq:invariant_derivatives}
\end{equation}

The same derivatives as in eq. (\ref{eq:invariant_derivatives}) apply to the the derivatives of the isochoric invariants with respect to $\bar{\mathbf{C}}$. However, when using the split into volumetric and isochoric parts, we need the projection

\begin{equation}
\frac{\partial \bar{\mathbf{C}}}{\partial \mathbf{C}} =\mathbb{P} = J^{-2/3}\left(\mathbb{I}-\frac{1}{3}\mathbf{C}\otimes \mathbf{C}^{-1}\right)\, .
\label{eq:P_projection}
\end{equation}

The tensor $\mathbb{I}$ in eq.(\ref{eq:P_projection}) denotes the fourth order identity. Bringing it all together, the second Piola-Kirchhoff stress takes the form 

\begin{equation}
\mathbf{S} = 2\frac{\partial \psi}{\partial \mathbf{C}} = 2\frac{\partial \psi}{\partial \bar{\mathbf{C}}}:\mathbb{P} +\mathbf{S}_{\mathrm{vol}}= \bar{\mathbf{S}}:\mathbb{P} + \mathbf{S}_{\mathrm{vol}}\, ,
\end{equation}

with

\begin{equation}
\bar{\mathbf{S}} = 2\frac{\partial \psi_1}{\partial \bar{I}_1} \mathbf{I} + 2\frac{\partial \psi_2}{\partial \bar{I}_2} (\bar{I}_1 \mathbf{I} - \bar{\mathbf{C}}^{-1}) + 2\frac{\partial \psi_{4a}}{\partial \bar{I}_{4a}} \mathbf{a}_0\otimes \mathbf{a}_0 + 2\frac{\partial \psi_{4s}}{\partial \bar{I}_{4s}} \mathbf{s}_0\otimes \mathbf{s}_0 \, .
\label{eq:Sbar}
\end{equation}

It is possible to extend eq. (\ref{eq:psi_polyconvex_additive1}) to capture even a wider class of materials. Convex linear combinations of the invariants maintain polyconvexity with respect to $\mathbf{F}$. Therefore, in addition to the invariants in eq.(\ref{eq:invariants}) or their isochoric counterparts eq.(\ref{eq:iso_invariants}), we can consider the mixed invariants 

\begin{equation}
K_{ij} =  \alpha_{ij}I_i+(1-\alpha_{ij})I_j \, ,
\end{equation}

and the corresponding isochoric versions $\bar{K}_{ij}$. The family of strain energies considering these mixed terms has the following structure

\begin{equation}
\psi = \sum \psi_{i}(\bar{I}_i) + \sum \psi_{ij}(\bar{K}_{ij}) +\psi_{vol}(J)\, . 
\label{eq:psi_polyconvex_additive2}
\end{equation}

The expression for $\bar{\mathbf{S}}$ in this more general case is analogous to eq. (\ref{eq:Sbar}) but with additional terms to account for the $\psi_{ij}$ contributions. 

\subsection*{Uniaxial, pure shear, and biaxial loading}
For the specific case of isotropic uniaxial deformation of a perfectly incompressible material, the deformation depends on the single stretch $\lambda$, and the stress in the direction of the applied stretch is 

\begin{equation}
P = 2(\lambda - \lambda^{-2}) \left(\frac{\partial \psi}{\partial I_1} +\frac{1}{\lambda}\frac{\partial \psi}{\partial I_2}\right)\, .
\label{eq:P_UTl}
\end{equation}

For pure shear deformation of a wide but thin specimen, the nominal stress in the direction of the applied stretch $\lambda$ is

\begin{equation}
P = 2(\lambda - \lambda^{-3}) \left(\frac{\partial \psi}{\partial I_1} +\frac{\partial \psi}{\partial I_2}\right)\, .
\label{eq:P_PS}
\end{equation}

The third loading case of interest for thin incompressible isotropic materials is equibiaxial tension defined by the stretch $\lambda$. For this loading, the nominal stress in the two principal directions of applied stretch is the same and equal to 

\begin{equation}
P = 2(\lambda - \lambda^{-5}) \left(\frac{\partial \psi}{\partial I_1} +\lambda^2\frac{\partial \psi}{\partial I_2}\right)\, .
\label{eq:P_ET}
\end{equation}

Lastly we consider an incompressible transversely anisotropic material under arbitrary biaxial loading specified by the two stretches $\lambda_x$, $\lambda_y$. Without loss of generality we set $\mathbf{a}_0=[1,0,0]$, $\mathbf{s}_0=[0,1,0]$. The in plane nominal stresses are 

\begin{equation}
\begin{aligned}
&P_{xx} =  \frac{\partial \psi}{\partial I_1} \lambda_x + \frac{\partial \psi}{\partial I_2} (I_1 \lambda_x - \lambda_x^3) + \frac{\partial \psi}{\partial I_{4a}} \lambda_x - p\lambda_x^{-1}\\
&P_{yy} =  \frac{\partial \psi}{\partial I_1} \lambda_y + \frac{\partial \psi}{\partial I_2} (I_1 \lambda_y - \lambda_y^3) + \frac{\partial \psi}{\partial I_{4s}} \lambda_y - p\lambda_y^{-1} \, ,
\end{aligned}
\label{eq:P_biax_aniso}
\end{equation}

with the pressure Lagrange multiplier solved from the plane stress condition

\begin{equation}
p = \frac{\partial \psi}{\partial I_1} \lambda_z^2 + \frac{\partial \psi}{\partial I_2} (I_1 \lambda_z^2 - \lambda_z^4)\, ,
\label{eq:p_biax}
\end{equation}

and the normal stretch obtained from the incompressibility constraint

\begin{equation}
\lambda_z = \frac{1}{\lambda_x \lambda_y}  \, .
\label{eq:lamz_biax}
\end{equation}

\subsection*{CANN models}

To construct convex non-decreasing functions to represent the energy in eq. (\ref{eq:psi_polyconvex_additive2}), one way is to borrow from the architecture of feed forward neural networks but using only convex non-decreasing activation functions on a polynomial expansion. The method is illustrated in Fig. \ref{fig01}a. Starting from $\mathbf{F}$, the invariants eq.(\ref{eq:invariants}) are computed in a pre-processing step. For ease of implementation and to improve the optimization step during model training,  consider the normalized invariants 

\begin{equation}
\hat{I}_i = (I_i - a_i)/b_i
\label{eq:hatI}
\end{equation}

where $a_1=a_2=3$, $a_{4a}=a_{4s}=1$, and $b_i$ is a normalizing constant such that the range of $\hat{I}_i$ is approximately $[0,3]$. Note that in the case of full incompressibility as assumed from now on, the normalized invariants strictly satisfy $\hat{I}_i\geq 0$. For compressible or nearly incompressible materials, simply replace $I_i$ with the isochoric counterpart $\bar{I}_i$ in eq.(\ref{eq:hatI}). For the mixed invariants, the normalized version is

\begin{equation}
\hat{K}_{ij} = \alpha_{ij}\hat{I}_i+(1-\alpha_{ij})\hat{I}_j\, ,
\label{eq:hatK}
\end{equation}

which also satisfies $\hat{K}_{ij} \geq 0$ as along as $\alpha_{ij}\in [0,1]$.
For the general case including anisotropy, the strain energy can be summarized as 

\begin{equation}
\begin{aligned}
\psi_{\mathrm{CANN}} &= \sum_{i,a,b} \psi_{i,ab} &+ &\sum_{i,j,a,b,i\neq j} \psi_{ij,ab} \\
&= \sum_{i,a,b} g_{i,ab} f_b(w_{i,ab} P_a(\hat{I}_{i})) &+  &\sum_{i,j,a,b,i\neq j} g_{ij,ab} f_b(w_{i,ab} P_a(\hat{K}_{ij}))\, ,
\end{aligned}
\end{equation}

where $P_a(x)=x^a$ is a basic polynomial expansion with $a\in\{1,2,3\}$ in our implementation, $f_b(\circ)$ is the activation function choice in our case either identity $f_1(x)=x$ or exponential $f_2(x)=\exp(x)-1$. The notation is the same for the mixed invariants. The weights $g_{i,ab}$, $w_{i,ab}$, $g_{ij,ab}$, $w_{ij,ab}$ are the trainable parameters of the model and need to be non-negative to maintain the convex non-decreasing output. 

For the rubber examples below, we only use the two main invariants $\hat{I}_1$, $\hat{I}_2$. For the skin example we have two ansatz. The simpler model includes contributions from $\hat{I}_1$, $\hat{I}_2$, $\hat{K}_{4a\,4s}$. The second option for the skin examples takes inputs $\hat{I}_1$, $\hat{I}_2$, $\hat{K}_{1\,2}$, $\hat{K}_{1\,4a}$, $\hat{K}_{1\,4s}$, $\hat{K}_{4a\,4s}$. 

Our choice of polynomials and activation functions guarantee the interpolation of convex non-decreasing functions of the inputs $\hat{I}_i$, $\hat{K}_{ij}$ in the domain $\hat{I}_i,\hat{K}_{ij}\geq 0$ provided that only non-negative weights are used which is easy to enforce. The non-negative condition on the domain, $\hat{I}_i,\hat{K}_{ij}\geq 0$, is trivially satisfied for incompressible materials, and satisfied for compressible or nearly incompressible materials if the isochoric invariants are used in eq. (\ref{eq:hatI}). Thus, CANNs \textit{a priori} result in polyconvex strain energy functions.

\subsection*{ICNN models}

This algorithm also relies on building convex functions of the normalized invariants and linear combinations of them. Let $X$ be the input to the first layer, and $\mathbf{Z}_{i-1}$ the output of layer $i-1$. Then, for layer $i$ the output is 

\begin{equation}
\mathbf{Z}_i = \mathrm{softplus}^2( \exp(\mathbf{W}^\top_{z,i}) \mathbf{Z}_{i-1}  + X \exp(\mathbf{W}_{x,i}) + \mathbf{b}_i ) \, , 
\end{equation}

parameterized by the weights $\mathbf{W}_{z,i}$, $\mathbf{W}_{x,i}$ and biases $\mathbf{b}_i$. For the first layer we have

\begin{equation}
\mathbf{Z}_1 = \mathrm{softplus}^2(  X \exp(\mathbf{W}_{x,1}) + \mathbf{b}_1 ) \, ,
\end{equation}

while for the last layer 

\begin{equation}
Z_n =  \exp(\mathbf{W}^\top_{z,n}) \mathbf{Z}_{n-1}  + X \exp(W_{x,n}) + \mathbf{b}_n  \, .
\end{equation}

This architecture retains convexity because $\mathrm{softplus}^2(\circ)$ is a convex non-decreasing function evaluated on linear combinations of the original input and the intermediate layer outputs using non-negative weights (enforced with the $\exp (\circ)$ function). Therefore, ICNNs can be used to create convex non-decreasing functions of the same normalized invariants $\hat{I}_i$ and normalized mixed invariants $\hat{K}_{ij}$ defined in eqs. (\ref{eq:hatI}),(\ref{eq:hatK}) for the CANN models. The general expansion is 

\begin{equation}
\psi_{\mathrm{ICNN}} = \sum_i \psi_i(\hat{I}_i) + \sum_{i,j,\, i\neq j} \psi_{ij}(\hat{K}_{ij}) \, .
\end{equation}

Similarly to the CANNs, for the rubber examples using ICNNs we only consider two functions $\psi_1(\hat{I}_1)$, $\psi_2(\hat{I}_2)$. For the anisotropic examples we have two models. The simpler one uses three functions $\psi_1(\hat{I}_1)$, $\psi_2(\hat{I}_2)$, $\psi_{4a\, 4s}(\hat{K}_{4a\, 4s})$. The second anisotropic model also includes the mixed terms $\psi_{1\, 4a}(\hat{K}_{1\, 4a})$, $\psi_{1\, 4s}(\hat{K}_{1\, 4s})$, $\psi_{1\, 2}(\hat{K}_{1\, 2})$.

\subsection*{NODE models}

In contrast to the two previous methods, NODEs avoid interpolation of the energy and interpolate the derivative functions directly. In the end, the derivatives with respect to the invariants are the ones that enter the definition of the stress, see eq. (\ref{eq:Sbar}). Consider the normalized invariant $\hat{I}_i$, the NODE is a feed-forward neural network with weights $\mathbf{W}$ and biases $\mathbf{b}$ that define the function $f(\circ)$ of the ODE

\begin{equation}
\frac{d y_i(\omega)}{d \omega}=f(y_i,\omega) \; , y_i(0)=\hat{I}_i \, ,
\label{eq:ODE}
\end{equation}

where $\omega$ is a pseudo-time auxiliary variable. The output of interest is the solution of the ODE at a fixed pseudo-time. In this implementation we choose $\omega=1$,

\begin{equation}
\frac{\partial \psi}{\partial \hat{I}_i} = y_i(1)  \, .
\end{equation}

Note that the output is directly the derivative of the strain energy. The key observation is that trajectories of ODEs do not intersect, thus for two initial conditions $y^{(a)}_i(0) \geq y^{(b)}_i(0)$, the ensuing trajectories continue to satisfy $y^{(a)}_i(\omega)\geq y^{(b)}_i(\omega)$. This implies

\begin{equation}
\left. \frac{\partial \psi}{\partial \hat{I}_i}\right|_{\hat{I}_1=\hat{I}_1^{(a)}}\geq \left. \frac{\partial \psi}{\partial \hat{I}_i}\right|_{\hat{I}_1=\hat{I}_1^{(b)}} \iff \hat{I}_1^{(a)}\geq \hat{I}_1^{(b)} \, .
\label{eq:monotonicity}
\end{equation} 

The monotonicity of the output eq.(\ref{eq:monotonicity}) is equivalent to convexity of the underlying $\psi$. For the mixed invariants, the NODE defines the derivative

\begin{equation}
\frac{\partial \psi}{\partial \hat{K}_{ij}} = y_{ij}(1)  \, ,
\end{equation}

for an ODE analogous to eq. (\ref{eq:ODE}). Therefore, when using NODE models we do not recover an analytical expression for $\psi_{\mathrm{NODE}}$. Nevertheless, the energy can be integrated if needed

\begin{equation}
\psi_{\mathrm{NODE}} = \sum_{i} \int_{\hat{I}_i} \frac{\partial \psi}{\partial \hat{I}_i} + \sum_{i,j \, i\neq j} \int_{\hat{K}_{ij}} \frac{\partial \psi}{\partial \hat{K}_{ij}} \, 
\end{equation}

along a given trajectory over $\hat{I}_i$, $\hat{K}_{ij}$. Even though convexity of $\psi_{\mathrm{NODE}}$ with respect to the invariants is ensured by eq.(\ref{eq:monotonicity}), to construct convex non-decreasing functions the additional restriction of zero biases $\mathbf{b}=0$ is applied. With this last correction, the energy $\psi_{\mathrm{NODE}}$ is automatically polyconvex.

\subsection*{Benchmark datasets and test cases}

We consider two datasets in this study, a classicall rubber dataset including uniaxial tension (UT), pure shear (PS), and equibiaxial tension (ET) nominal stress-stretch data from \cite{linka2023new}. The other dataset is from porcine skin and consists of three biaxial tests: strip biaxial in the $x$ direction (SX), i.e. $\lambda_x=\lambda$ is applied and the orthogonal direction is kept at $\lambda_y=1$, strip biaxial in $y$ direction (SY), and equibiaxial tension (EB). Data from the skin data comes from \cite{tac2022data}, and is also nominal stress-stretch data. 

\section*{Results}

\subsection*{Performance on rubber dataset}

\begin{figure*}[h!]
\centering
\includegraphics{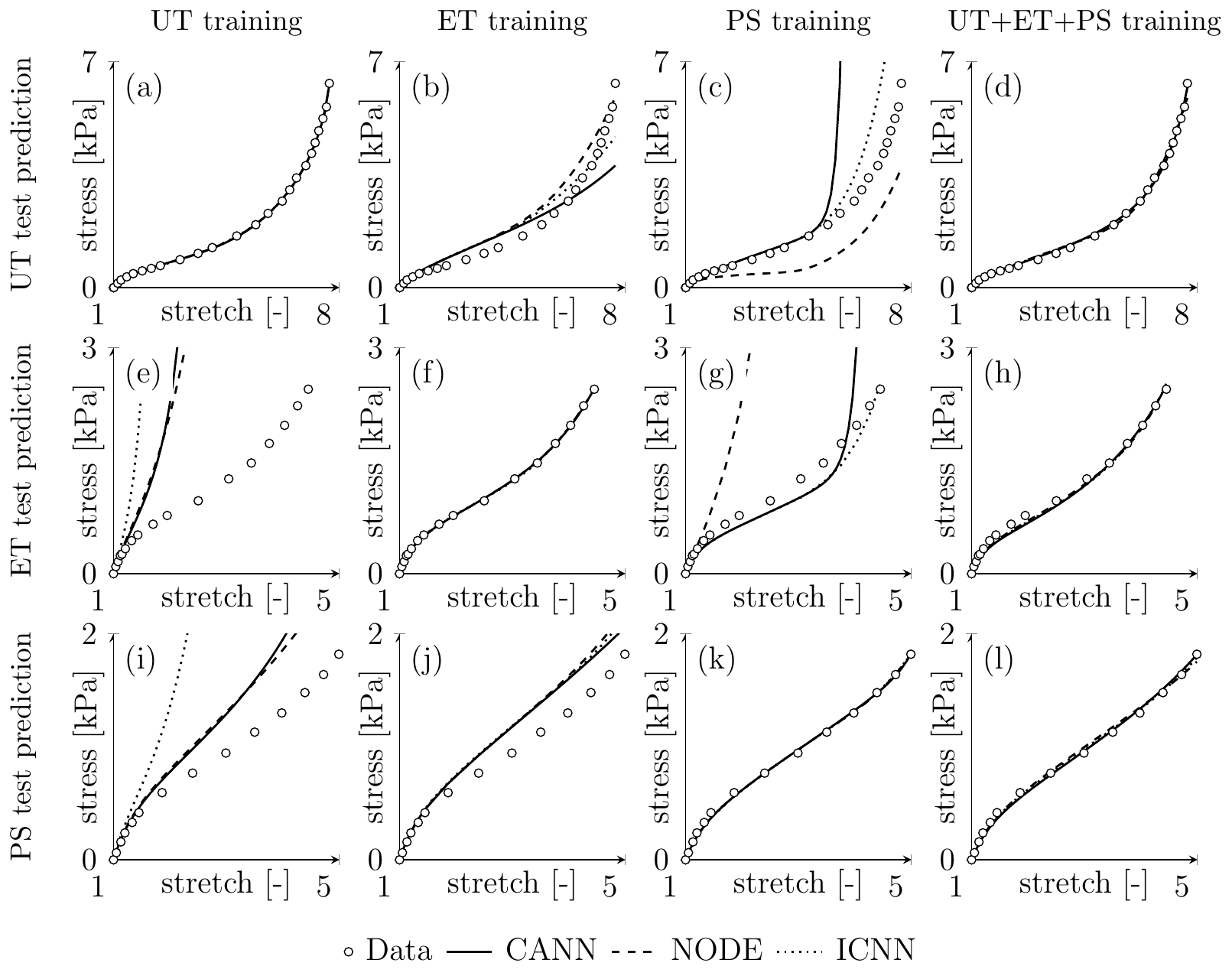} \vspace{-4mm}
\caption{Performance of CANN, NODE and ICNN moddels on rubber nominal stress-stretch data. Trained on uniaxial tension (UT), models are compared against UT data (a) but also against equibiaxial tension (ET) data (e) and pure shear (PS) data (i). Trained on ET, models are evaluated on UT, ET, PS data (b,f,j) respectively. Similarly, trained on PS data, comparison against UT, ET, PS curves (c,g,k). All three models can capture all three datasets when trained on all data simultaneously (d,h,l). }
\label{fig02} 
\end{figure*}

The rubber dataset contains three mechanical tests as described in the Methods Section. To test the ability of the data-driven methods to extrapolate we trained first against one of the three tests and compared against the other two. Results are depicted in the first three columns of Fig. \ref{fig02}. Not surprisingly, all three methods perfectly capture the loading curve on which they are trained on (Fig \ref{fig02}a,f,k). However, the methods have difficulty extrapolating. Depending on which test was used for training, the performance on the validation data varies. When trained on uniaxial data, predictions on the other two tests are inaccurate, with stiffer predictions in all cases compared to the data (Fig \ref{fig02}e,i). The ICNN trained on pure-shear data is still able to capture the response in biaxial and uniaxial loading (Fig \ref{fig02}c,g). In contrast, the CANN model trained on PS data can predict UT and ET data up to an intermediate stretch after which the prediction exponential increases and diverges from the data. The NODE trained on PS data performs well on the UT dataset but not on the ET dataset. Equibiaxial training appears to be the best for extrapolating for all three methods. The prediction for ET data matches closely the experiments, see Fig.\ref{fig02}f, and the predictions for uniaxial and pure shear qualitatively match the observed response albeit with some error (Fig.\ref{fig02}b,j). To verify that the methods are indeed able to capture the entire response of the material, the last column of Fig. \ref{fig02} shows predictions when CANN, NODE, ICNN models are trained on all data at once. All three methods flawlessly interpolate the entire dataset (Fig.\ref{fig02}d,h,l).

Results in Fig. \ref{fig02} are representative, yet, they correspond to single fit from the CANN, NODE and ICNN models. To show the robust performance of the data-driven methods, we repeat the training 50 times and compute $R^2$ values for each trained model. The $R^2$ values are shown in Fig. \ref{fig03} in a layout analogous to the representative training Fig. \ref{fig02}. The $R^2$ values confirm the previous observations from Fig. \ref{fig02}. For uniaxial training, $R^2$ values on UT data are approximately one always (Fig. \ref{fig03}a) but there is little predictive performance on the other two tests (Fig. \ref{fig03}e,i). For PS training we confirm that NODE and ICNN are able to capture the PS response (Fig. \ref{fig03}k), the UT response (Fig. \ref{fig03}c), but not the ET data (Fig. \ref{fig03}g). CANN models can capture the pure shear response just as well (Fig. \ref{fig03}k), but unable to extrapolate to the other two loading cases (Fig. \ref{fig03}c,g). With 50 instances of model fitting,  we can confidently state that equibiaxial tests are indeed the ones that allow the three machine learning models to better extrapolate to other loading cases. $R^2$ values in Fig. \ref{fig03}b,f,j are always greater than $0.656$, with narrow standard deviations. Fig. \ref{fig03}d,h,l also confirms that when trained on all data at once, CANN, ICNN and NODE have no trouble fitting the data, achieving $R^2$ on average $0.971$, $0.997$, $0.997$ for each of the methods respectively. 

\begin{figure*}[h!]
\centering
\includegraphics{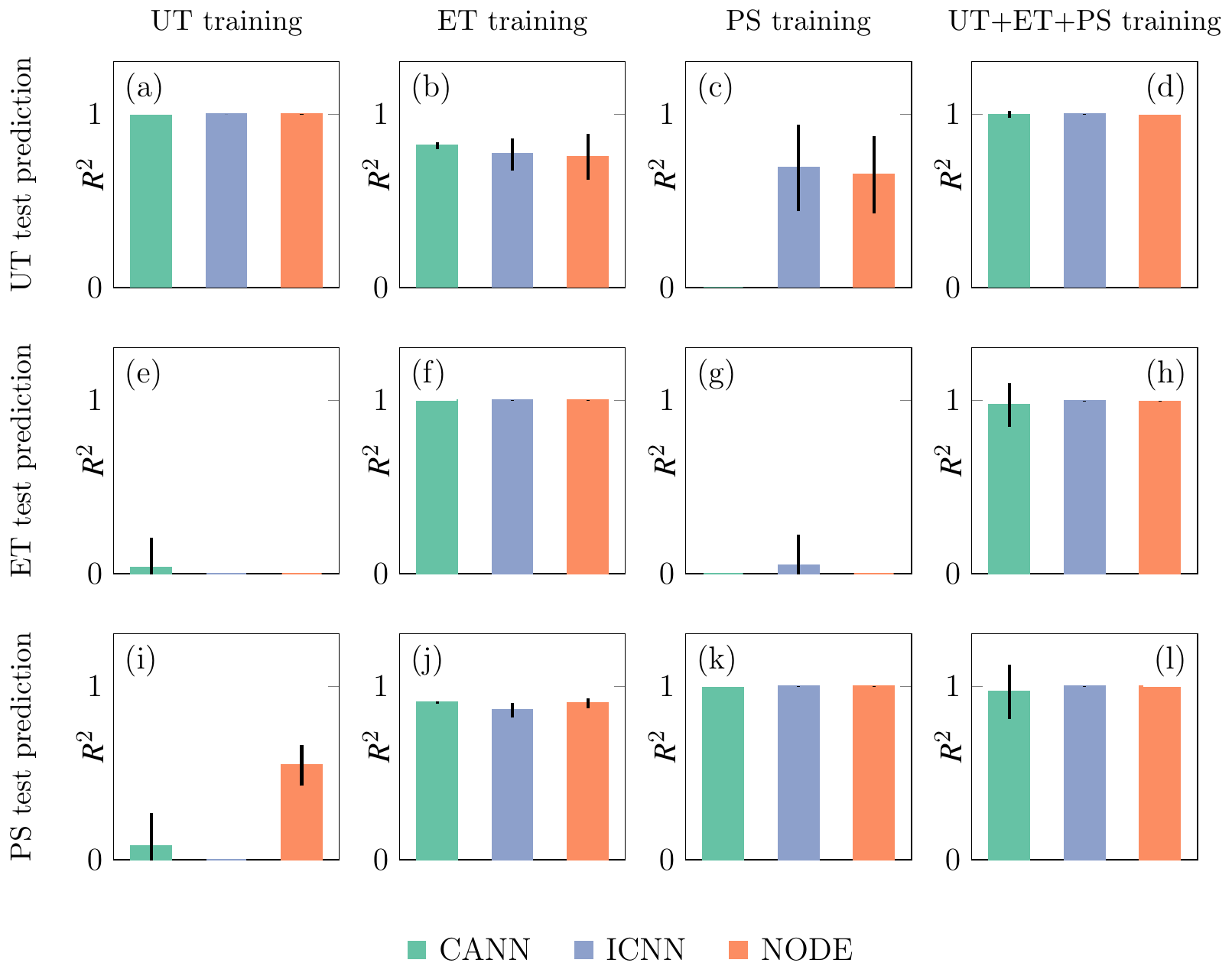} \vspace{-4mm}
\caption{Performance of the data-driven constitutive models in terms of average $R^2$ values from XX training with different initialization. CANN, NODE and ICNN moddels trained on uniaxial tension (UT) are compared against UT data (a), equibiaxial tension (ET) data (e), and pure shear (PS) data (i). Trained on ET, models are evaluated on UT, ET, PS data (b,f,j) respectively. Trained on PS data, comparison is done against UT, ET, PS curves (c,g,k). All three models have $R^2$ values near one for all data when trained on all data simultaneously (d,h,l).}
\label{fig04} 
\end{figure*}

\subsection*{Performance on skin dataset}

The anisotropy of skin leads to poorer capacity of the three algorithms for extrapolation. Trained with either strip biaxial in $x$, strip biaxial $y$, or equibiaxial tension, the three methods can capture the response they are trained on but unable to extrapolate, as illustrated in the first three columns of Fig. \ref{fig04}. To capture the transversely anisotropic response of skin, the number of parameters and flexibility of the functional space available to the three data-driven methods enables them to produce complex response, but at the same time it leads to unconstrained and poor predictions outside of the training region. For instance, trained on SX data, predictions under SX loading are remarkably accurate (Fig. \ref{fig04}a), but CANN models tend to predict stiffer responses in EB and SY loading (Fig. \ref{fig04}e,i); NODE predics stiffer response in SY loading (Fig. \ref{fig04}i) but accurate response in EB loading (Fig. \ref{fig04}e), and ICNN performs well in EB loading (Fig. \ref{fig04}e) but predicts soft response compared to the data in SY loading (Fig. \ref{fig04}i). To verify if the models are able to capture the entire dataset we trained CANN, NODE, ICNN modes with all the data simultaneously and show the fits in Fig. \ref{fig04}d,h,l. All three methods can capture the response when trained on all data, however, fits are not perfect compared to the individual test fitting in Fig. \ref{fig04}a,f,k. The poorer performance in the simultaneous fitting is consistent between all three methods and suggests that the data themselves might be inconsistent with the assumption of hyperelasticity, that there are experimental errors, or that the functional space available to the data-driven models needs to be even richer.

\begin{figure*}[h!]
\centering
\includegraphics{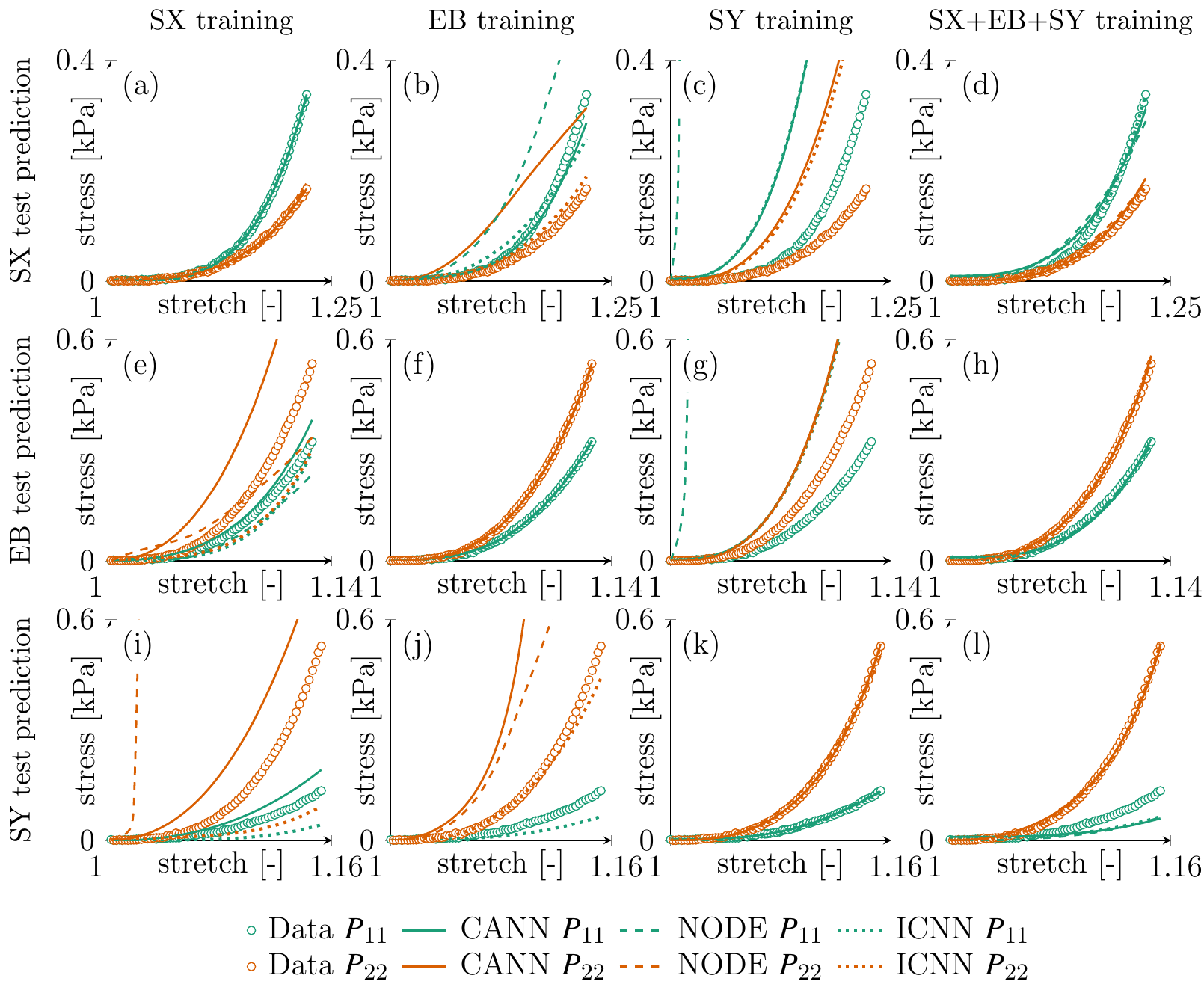} \vspace{-4mm}
\caption{Performance of CANN, NODE and ICNN models against the skin mechanics dataset. Trained on strip $x$ (SX) data, the three models were compared to SX (a), equibiaxial (EB) (e) and strip $y$ (SY) data (i). Trained on EB data, comparison to SX, EB, SY data is shown in (b,f,j). Trained on SY data, comparison to SX, EB, SY is shoin in (c,g,k). Trained on all data simultaneously, the three methods can capture SX (d), EB(h) and SY response (l). }
\label{fig03} 
\end{figure*}

\begin{figure*}[h!]
\centering
\includegraphics{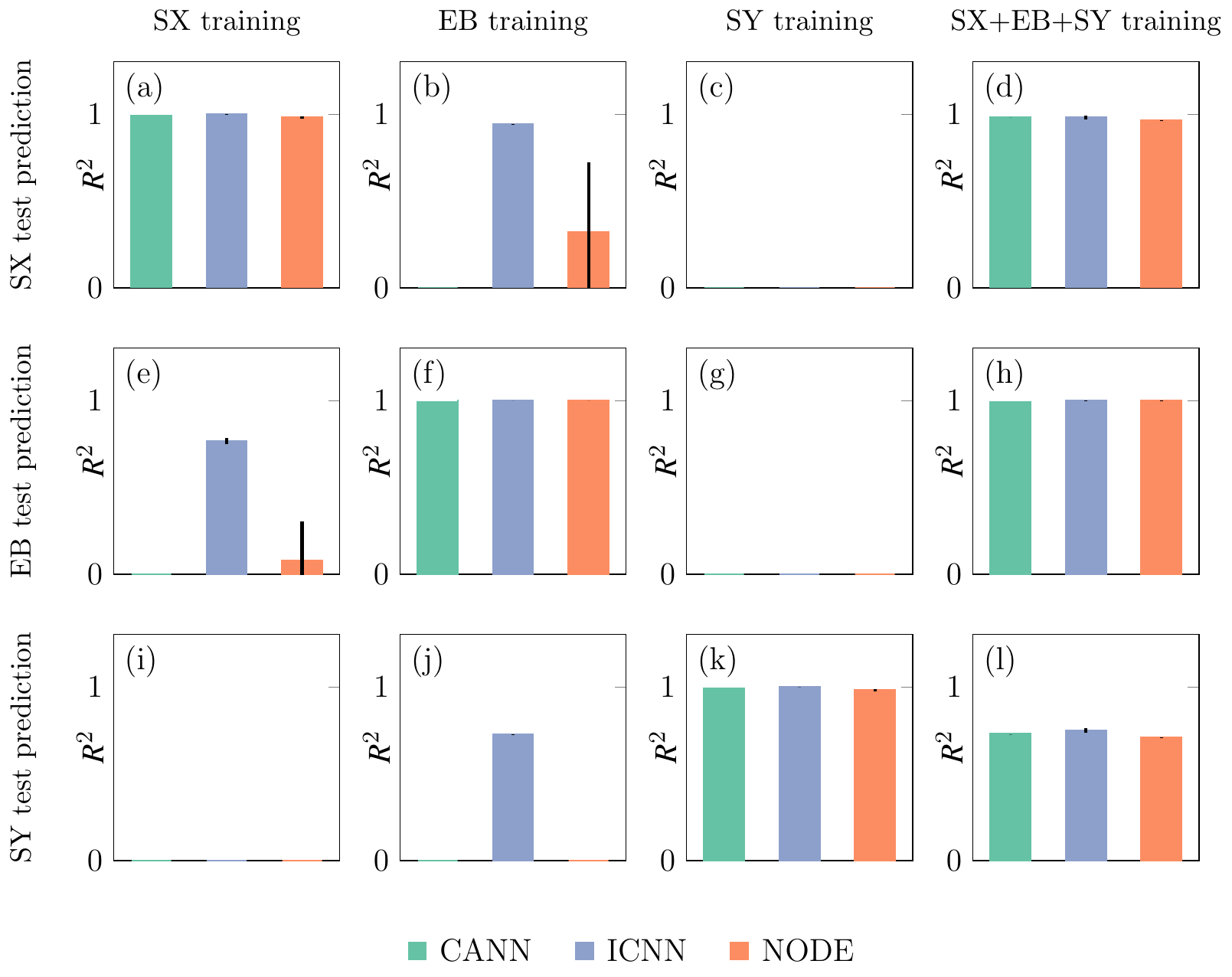} \vspace{-4mm}
\caption{Performance of the data-driven methods on the skin dataset in terms of $R^2$ scores for data from three type of tests: strip biaxial $x$ (SX), equibiaxial (EB), strip biaxial in $y$ direction (SY). Trained on SX data (a), the models are tested agains EB (e) and SY (i) data. Trained on EB data (f), the models are tested on SX (b) and SY (j) loading. When trained on SY data (k), the models cannot capture SX (c) or EB (g) response.  When trained on all data simultaneously, CANN, ICNN, and NODE models can capture all three types of loading: SX (d), EB (h), SY (l). }
\label{fig05} 
\end{figure*}

A more quantitative analysis of the performance is reported in Fig. \ref{fig05}, which shows $R^2$ values computed after $10$ instances of model training with different, random initialization. Just as observed in the representative fits of Fig. \ref{fig04}, the $R^2$ scores on the loading used for training are near one (Fig. \ref{fig05}a,f,k), but they are low or even near zero for the validation cases (Fig. \ref{fig05}b,c,e,g,i,j). Surprisingly, there is still some information from the equibiaxial test (Fig. \ref{fig05}f) that is useful for extrapolation to the strip biaxial loading cases (Fig. \ref{fig05}b,j). Training on the strip biaxial tests, SY has no information for the SX or EB data (Fig. \ref{fig05}c,g), whereas SX training does lead to some $R^2>0$ for EB (Fig. \ref{fig05}e) but not SY data (Fig. \ref{fig05}i). The methods are able to consistently interpolate the entire data from all three tests regardless of random initialization (Fig. \ref{fig05}d,h,l). 

\subsection*{Regularity of second derivatives}

Thus far we have focused on the performance of the data-driven models to capture stress-stretch data, which directly relates to strain energy derivatives. However, using these highly nonlinear model in large scale physics solvers, either implicit dynamics or equilibrium, requires computation of second derivatives of the energy. Therefore, even though second derivatives are not related to any data, we are interested in the regularity of the second derivatives for CANN, ICNN and NODE models. 

\begin{figure*}[h!]
\centering
\includegraphics{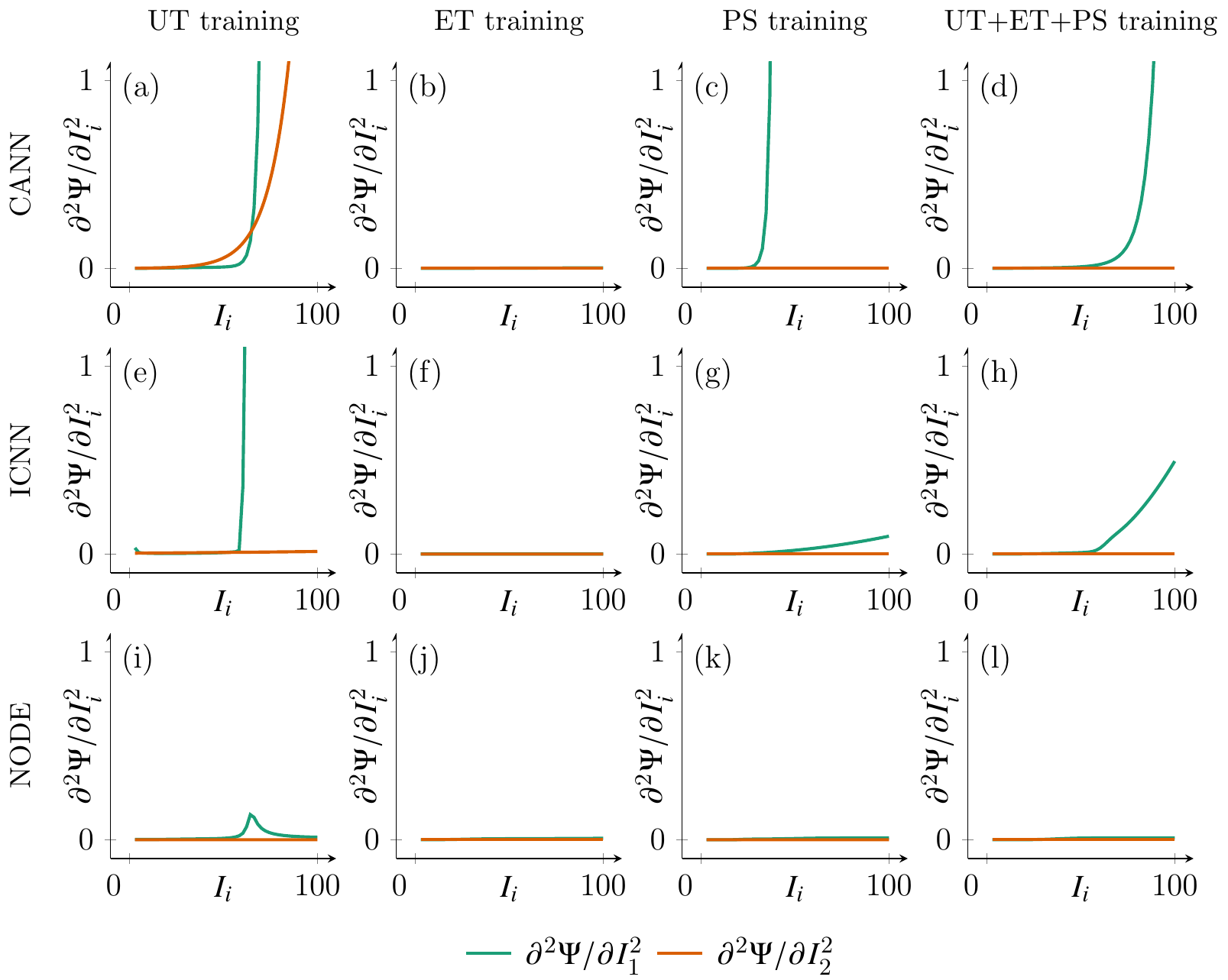} \vspace{-4mm}
\caption{Second derivatives of strain energy functions  predicted by the data-driven models trained with rubber data. Training with UT only, predictions are done for CANN (a), ICNN (e) and NODE (i) models. Similarly, second derivatives for the three methods are shown for ET training only (b,f,j), PS training only (c,g,k) and tained against all data (d,h,l). }
\label{fig06} 
\end{figure*}

For the rubber benchmark the models are based on the interpolation of two functions $\partial \psi(I_1)/\partial I_1$, $\partial \psi(I_2)/\partial I_2$. Fig. \ref{fig06} shows the second derivatives $\partial^2 \psi(I_1)/\partial I_1^2$, $\partial^2 \psi(I_2)/\partial I_2^2$ with the same layout as Fig. \ref{fig02} and Fig. \ref{fig03}. It is surprising that even though all three methods capture the stress data quite well, they differ substantially in terms of their second derivatives. This reflects that there are many strain energy functions $\psi(I_1,I_2)$ that are polyconvex and that can capture the stress-stretch data under uniaxial, pure shear, and equibiaxial loading. The CANN, ICNN and NODE are suited to capture different functions within the large space of functions available to each method. The consistent trend in Fig. \ref{fig06} is that the CANN models lead often to exponential second derivatives because one of the two key activation functions is the exponential. In contrast, the NODE model is the one with the smallest second derivatives in all cases. For all the three methods, the second derivatives are smooth functions.

\begin{figure*}[h!]
\centering
\includegraphics{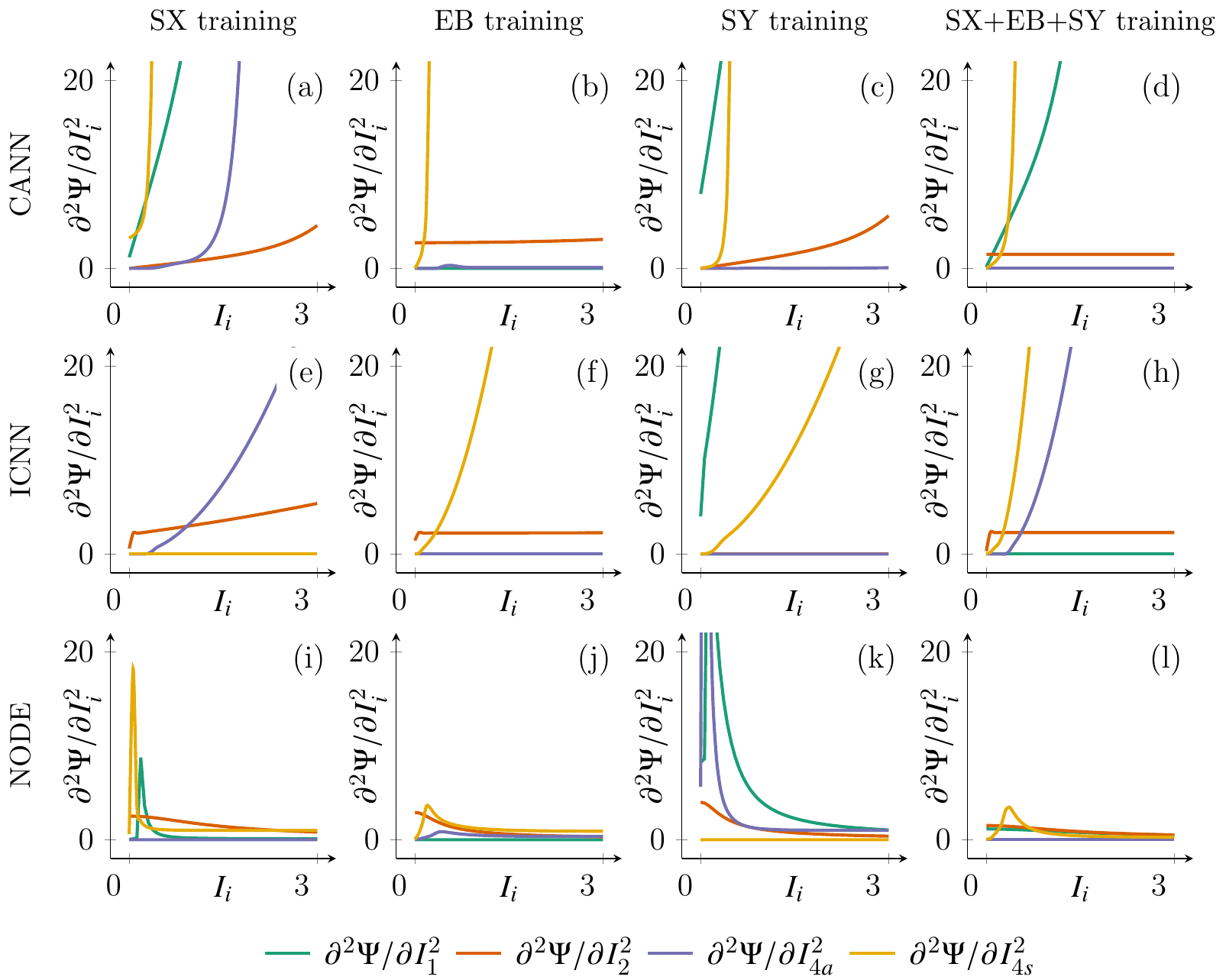} \vspace{-4mm}
\caption{Second derivatives of strain energy when trained with skin data. CANN predictions (a-d), ICNN predictions (e-h), and NODE predictions (d-l). Columns correspond to the data used in training. First three columns correspond to either SX, EB, or SY data only. Last column shows predictions when models are trained on all data simultaneously. }
\label{fig07} 
\end{figure*}

For the skin benchmark, there are more functions being interpolated by the three data-driven frameworks. As a result, Fig. \ref{fig07} shows the second derivatives $\partial^2 \psi/\partial I_1^2$, $\partial^2 \psi/\partial I_2^2$, $\partial^2 \psi/\partial I_{4a}^2$, and $\partial^2 \psi/\partial I_{4b}^2$. Consistent with the rubber data NODE second derivatives are the smallest out of the three methods. The second derivatives might increase for some initial range of deformation but tend to smaller values toward the end of the testing ranges (Fig. \ref{fig07}i-l). The CANN (Fig. \ref{fig07}a-d) and ICNN methods ((Fig. \ref{fig07}e-h) have increasing second derivatives over the range of the invariants. Also similar to the rubber benchmark, here we see that even though all three methods perform similarly on the stress-stretch predictions (see Fig. \ref{fig04}), they do so by interpolating different functions $\psi(I_1,I_2,I_{4a},I_{4s})$.   

\subsection*{Model efficiency}

A key question and common criticism of data-driven models, particularly neural network-based models, is that increasing the number of trainable parameters logically allows the methods to capture the limited data better and better, but at the risk of over-fitting. The polyconvexity constraint, enforced exactly for CANN, ICNN and NODE models, prevents nonphysical extrapolation, much like expert models. On the other hand, expert models and some non-parametric data-driven methods \cite{fuhgPhysicsinformedDatadrivenIsotropic2022} are generally very efficient and capture the data reasonably well with very few parameters. We test how efficiently can the data-driven models interpolate the data, i.e. we ask how does the error decrease as a function of the number of trainable parameters. 

\begin{figure*}[h!]
\centering
\includegraphics{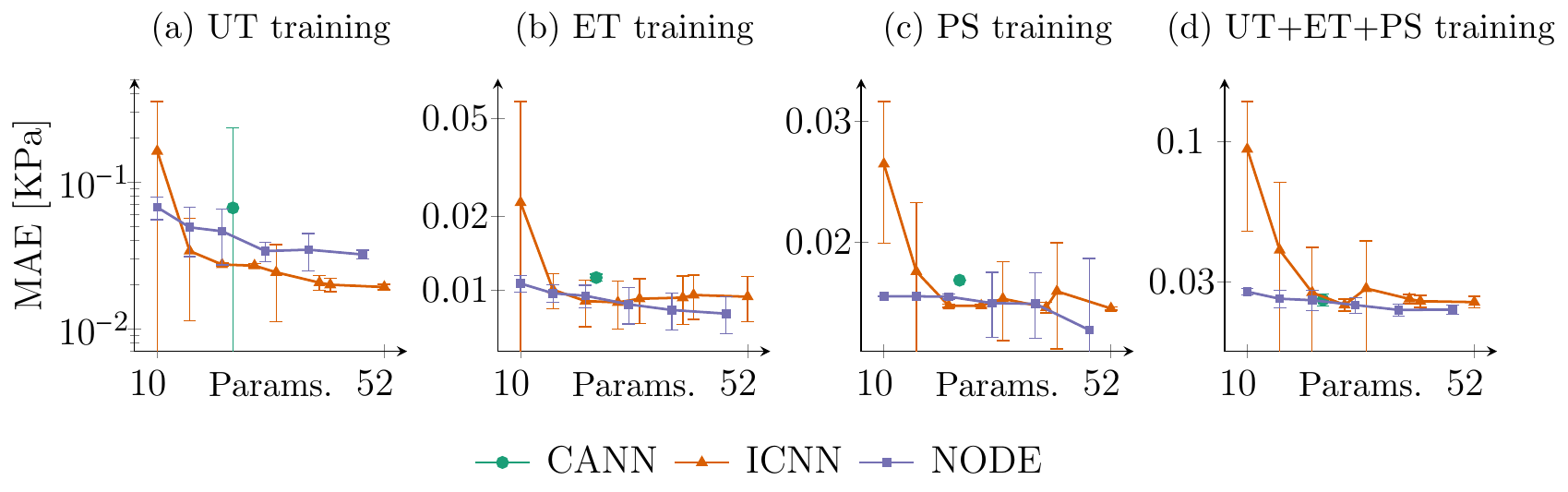} \vspace{-4mm}
\caption{Model efficiency for the rubber dataset depicted in terms of mean absolute error (MAE) against number of trainable parameters. Columns correspond to the type of data used for training: UT (a), ET (b), PS (c), or all data used simultaneously during training (d). Note that for the CANN model the number of parameters is fixed and a single point is shown for the CANN model in each panel. The ICNN and NODE are neural network-based models and the number of trainable parameters increases with number of layers and layer depth. }
\label{fig08} 
\end{figure*}

Fig. \ref{fig08} shows the efficiency plots for the rubber benchmark. The structure of the CANN model is between that of a neural network and an expert model. As a result, there is a single point for the CANN model for each of the plots in Fig. \ref{fig08}. For ICNN and NODE models, the error decays with increasing number of parameters, as expected. When there are 52 trainable parameters, the NODE and ICNN show similar performance in all the training cases. However, the drop in the error is more pronounced for the ICNN compared to the NODE framework. This suggests that the NODE model can capture the data well even with very few parameters.

\begin{figure*}[h!]
\centering
\includegraphics{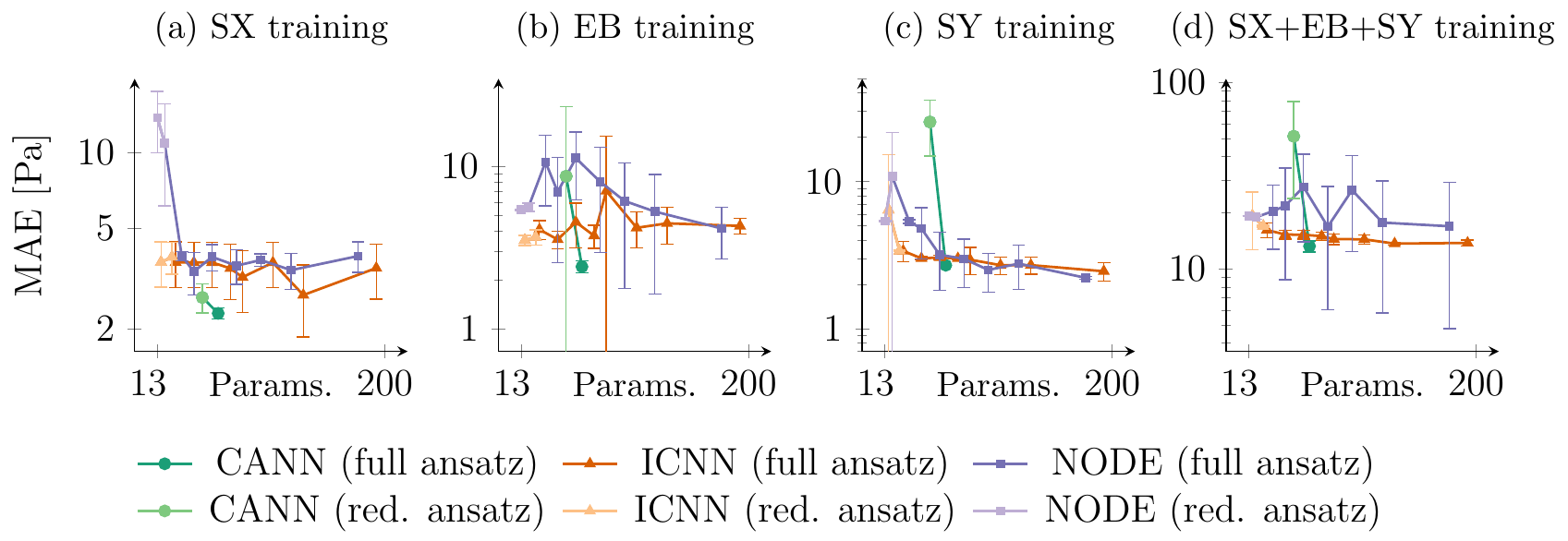} \vspace{-4mm}
\caption{CANN, ICNN and NODE model efficiency for the skin benchmark shown as mean absolute error (MAE) against number of trainable parameters and model complexity. For each of the three models, two ansatz are used: a reduced expansion based on ((\ref{eq:psi_polyconvex_additive1}), or a full expansion (\ref{eq:psi_polyconvex_additive12}). Plots show efficiency corresponding to different training cases: SX (a), EB (b), SY (c), or all data simultaneously (d).  }
\label{fig09} 
\end{figure*}

The efficiency trends are not preserved for the anisotropic skin data as shown in Fig. \ref{fig09}. In this case, in order to explore the effect of the number of parameters on the accuracy of the methods we follow two strategies: reducing the ansatz by interpolating only the functions in (\ref{eq:psi_polyconvex_additive1}), or using the full expansion (\ref{eq:psi_polyconvex_additive2}) but changing the number of trainable parameters. For the CANN model, which has a fixed number of parameters when considering either (\ref{eq:psi_polyconvex_additive1}) or (\ref{eq:psi_polyconvex_additive2}), we observe that the full ansatz has lower errors than the reduced one for all training cases. The flexibility of the framework increases by including the mixed terms, which helps with capturing the data better. This is consistent with the development of mixed invariant terms in popular closed-form constitutive equations such as the Gasser-Ogden-Holzapfel model \cite{gasser2005GOH}. The ICNN and NODE also show decreasing errors when going from the reduced model (\ref{eq:psi_polyconvex_additive1}) to the model including mixed invariants (\ref{eq:psi_polyconvex_additive2}). The improvement is much more pronounced for the NODE model compared to the ICNN one. In contrast to the rubber dataset, for skin, increasing the number of parameters of the neural networks used in the NODE models leads to a large decrease in error. The ICNN model error decreases only slightly with increasing number of parameters. At the upper end of the range considered, i.e. approaching 200 parameters, both ICNN and NODE perform similarly. The most efficient of the methods for skin data is the CANN, which achieves the lowest errors with the lowest number of parameters.  

\section*{Discussion}

This manuscript analyzes the performance of three data-driven methods for isotropic and anisotropic hyperelastic materials that automatically satisfy objectivity and polyconvexity of the strain energy. Traditional closed-form models rely on selecting few functional terms to capture the response of a material in a parsimonious way with few parameters to fit. Closed-form expressions are an elegant solution but also have the major downside of sacrificing accuracy. Data-driven methods have the flexibility to perfectly interpolate data. However, a paucity in their adoption is the difficulty to guarantee basic physics constraints that are front and center in the design of expert models \cite{ehret2007polyconvex,fuhgModularMachineLearningbased2022}. Objectivity and polyconvexity are the key requirements to represent realistic materials. The CANN, ICNN, and NODE models studied here are constructed in such a way that they \textit{a priori} satisfy these essential physics constraints. Therefore, these three methods have the potential to revolutionize modeling and simulation of highly nonlinear materials. We show that the three methods can interpolate rubber and skin benchmark datasets for isotropic and anisotropic hyperelasticity. They capture the data almost perfectly and have some capacity to extrapolate when trained on part of the data. The second derivatives are smooth, which is needed for equilibrium and implicit dynamic solvers. The models show the expected trade-off between accuracy and number of parameters. Overall, either of the three modeling frameworks is suitable  for fully data-driven material modeling. 

The methods we analyze here stand in contrast to other recent developments in data-driven computational mechanics. The most obvious way of leveraging machine learning tools is to directly interpolate strain-stress data. There are methods along those lines developed in recent years \cite{kirchdoerfer2016data}. One limitation of these approaches is the inability to extrapolate.  
Another problem of dealing with stress data is that objectivity and polyconvexity are not satisfied \textit{a priori}. 
Data-driven models that capture the strain energy function are more similar to expert models \cite{liu2020}. CANN, ICNN and NODE fall on this category. For the data-driven models that interpolate the strain energy, one option to impose polyconvexity is through the loss function. These methods have had some success but have to carefully balance between imposing the constraint or achieving a higher accuracy \cite{tac2022data, ghaderiPhysicsInformedAssemblyFeedForward2020a}. Another approach is to select the best model out of a wide library of available models \cite{flaschel2023automated}. CANN, ICNN and NODE models have been recently developed to automatically satisfy polyconvexity which is a sufficient condition for the solution of boundary value problems in hyperelasticty. The original version of these methods was introduced in recent publications \cite{klein2022polyconvex,tac2021automatically,linka2023new}. However, benchmarking of the original formulations is challenging because different expansions of the energy were used in each case. In this work we have re-formulated the methods such that the same invariants and energy terms are used consistently across the three models. With this implementation we show that, because the constraints are embedded in the methods, there is no trade-off between model accuracy and enforcing the physics. The three methods can get accurate representation of the data, show positive second derivatives of the energy with respect to the invariants, and perform robustly even with random initialization. 

The ideas behind each method are different and this translates into slight differences in their performance. CANN leverages the structure of feed-forward neural networks but uses a fixed number of available terms. Fitting a CANN involves finding the weights of the different terms. As a result, CANNs produce parsimonious models but are inherently limited by the number of functional terms. Despite the fixed structure, CANNS perform well on the benchmark datasets of this study. ICNNs rely on building convex functions by using nested linear combinations of convex non-decreasing functions in every layer of an otherwise conventional feed-forward neural network structure. NODEs deal directly with the energy derivative functions and  leverage monotonicity of ODEs to get monotonic derivative functions (which implies convex functions). Because ICNNs and NODEs have an inner structure that resembles standard neural networks, they have  more freedom to adjust the number parameters by changing the number of layers or the depth within a layer. The efficiency plots reflect the trade-off between accuracy and number of parameters. As the number of parameters increases, the difference between NODE and ICNN models vanishes. Thus, all three methods can accurately and efficiently capture the data. 

The other notable difference between the methods is in the prediction of second derivatives of the strain energy. CANN and ICNN models tend to predict increasing second derivative functions. The NODE, in contrast with the other two methods, tends to predict smaller or vanishing second derivatives towards the end of the training region. This difference likely stems from the fact that ODEs have fixed points. In other words, the derivative predictions converge to a single value, consequently producing vanishing second derivatives. In all cases the second derivatives are smooth functions which is ideal for equilibrium and implicit dynamic solvers \cite{TAC2022115248}. This is in contrast with other data-driven methods that require additional regularization of the derivatives \cite{Vlassis202elastoplast}. Another strategy to work with the derivatives of the energy but avoid solving an ODE would be to explore integrable neural networks  \cite{teichert2019machine}.

The methods we benchmark here are have been designed to capture hyperelasticity. Even within the context of hyperelastic materials, the expansion of the energy can be done in different ways, potentially giving access to even wider classes of behavior \cite{steigmann2003isotropic}. More complex material response beyond hyperelasticity can also benefit from the flexibility of data-driven methods. There is still a gap in the development of physics-informed machine learning methods for dissipative materials such as plasticity and viscoelasticity \cite{zhang2022effects}. There have been data-driven methods in this direction, but without a complete set of built-in physics constraints \cite{xuLearningViscoelasticityModels2021, chenRecurrentNeuralNetworks2021}. Therefore, this is a central area for future work that can leverage the three existing frameworks reviewed and refined here. A second extension that is needed is modeling uncertainty in the material response. This is particularly relevant to biological tissue \cite{lee2020}. Neural network-based frameworks can capture perfectly the response of a single material and can easily retrained with new data, but a fully Bayesian approach would allow deeper understanding of population distribution. For example, it would allow us to model how skin properties change with age, sex, or ethnicity. A Bayesian framework would also allow monitoring of epistemic uncertainty to guide data collection and produce trustworthy simulations. The third item we want to highlight is the extension to multi-modality data. The three methods we explore are still based on stress-stretch data. In contrast, some expert models are built around the idea of microstructure modeling, multiscale simulations, or micromechanics arguments \cite{gasser2005GOH}. These ideas have started to permeate intro data-driven modeling \cite{holzapfel2021predictive,leng2021}. Alternatively, inferring material behavior from full-field displacements and global force data without relying on stress-stretch pairs has also gained attention recently \cite{flaschelUnsupervisedDiscoveryInterpretable2021,chen2021learning}. Physics-informed machine learning methods that can build on CANN, ICNN or NODE frameworks but also leverage images of the tissue microstructure or information about material composition are a natural next step. 


\section*{Conclusions}

We present three fully data-driven and physics-constrained methods for nonlinear material modeling: CANNs, ICNNs,  and NODEs. The methods capture hyperelastic material response perfectly on benchmark datasets of rubber and skin under three different loading cases. Evaluating their capacity to extrapolate, their efficiency, and the regularity of their second derivatives, we conclude that even though the methods have different features, they all have comparable low errors which decay with parameter and model complexity, have smooth second derivatives, and have some capacity to extrapolate. In summary, these methods hold the key for high-fidelity modeling of arbitrary material behavior without the need to select closed-form expressions. Code and data are available with this submission and we are confident that these resources complement our detailed analysis and will favor the ongoing development and refinement of data-driven computational mechanics. 

\section*{Acknowledgements}

This work was supported by the National Institute of Arthritis and Musculoskeletal and Skin Diseases, National Institutes of Health, United States under award R01AR074525.

\section*{Supplementary information}

The code associated with this publication is available at \url{https://github.com/tajtac/hypermodelcomp}.


\begin{thebibliography}{51}
\expandafter\ifx\csname natexlab\endcsname\relax\def\natexlab#1{#1}\fi
\providecommand{\bibinfo}[2]{#2}
\ifx\xfnm\relax \def\xfnm[#1]{\unskip,\space#1}\fi
\bibitem[{Lee et~al.(2018)Lee, Turin, Gosain, Bilionis, and Tepole}]{lee2020}
\bibinfo{author}{T.~Lee}, \bibinfo{author}{S.~Y. Turin}, \bibinfo{author}{A.~K.
  Gosain}, \bibinfo{author}{I.~Bilionis}, \bibinfo{author}{A.~B. Tepole},
\newblock \bibinfo{title}{Propagation of material behavior uncertainty in a
  nonlinear finite element model of reconstructive surgery},
\newblock \bibinfo{journal}{Biomechanics and Modeling in Mechanobiology}
  \bibinfo{volume}{17} (\bibinfo{year}{2018}) \bibinfo{pages}{1857--1873}.
\bibitem[{Duriez and Bieze(2017)}]{duriez2017soft}
\bibinfo{author}{C.~Duriez}, \bibinfo{author}{T.~Bieze},
\newblock \bibinfo{title}{Soft robot modeling, simulation and control in
  real-time},
\newblock in: \bibinfo{booktitle}{Soft Robotics: Trends, Applications and
  Challenges}, \bibinfo{publisher}{Springer}, \bibinfo{year}{2017}, pp.
  \bibinfo{pages}{103--109}.
\bibitem[{Limbert(2019)}]{limbert2019skin}
\bibinfo{author}{G.~Limbert}, \bibinfo{title}{Skin Biophysics: From
  Experimental Characterisation to Advanced Modelling},
  volume~\bibinfo{volume}{22}, \bibinfo{publisher}{Springer},
  \bibinfo{year}{2019}.
\bibitem[{Leshno et~al.(1993)Leshno, Lin, Pinkus, and
  Schocken}]{leshno1993multilayer}
\bibinfo{author}{M.~Leshno}, \bibinfo{author}{V.~Y. Lin},
  \bibinfo{author}{A.~Pinkus}, \bibinfo{author}{S.~Schocken},
\newblock \bibinfo{title}{Multilayer feedforward networks with a nonpolynomial
  activation function can approximate any function},
\newblock \bibinfo{journal}{Neural networks} \bibinfo{volume}{6}
  (\bibinfo{year}{1993}) \bibinfo{pages}{861--867}.
\bibitem[{Peng et~al.(2020)Peng, Alber, Tepole, Cannon, De, Dura-Bernal,
  Garikipati, Karniadakis, Lytton, Perdikaris et~al.}]{peng2020multiscale}
\bibinfo{author}{G.~C. Peng}, \bibinfo{author}{M.~Alber},
  \bibinfo{author}{A.~B. Tepole}, \bibinfo{author}{W.~R. Cannon},
  \bibinfo{author}{S.~De}, \bibinfo{author}{S.~Dura-Bernal},
  \bibinfo{author}{K.~Garikipati}, \bibinfo{author}{G.~Karniadakis},
  \bibinfo{author}{W.~W. Lytton}, \bibinfo{author}{P.~Perdikaris}, et~al.,
\newblock \bibinfo{title}{Multiscale modeling meets machine learning: What can
  we learn?},
\newblock \bibinfo{journal}{Archives of Computational Methods in Engineering}
  (\bibinfo{year}{2020}) \bibinfo{pages}{1--21}.
\bibitem[{Marsden and Hughes(1994)}]{marsden1994mathematical}
\bibinfo{author}{J.~E. Marsden}, \bibinfo{author}{T.~J. Hughes},
  \bibinfo{title}{Mathematical foundations of elasticity},
  \bibinfo{publisher}{Courier Corporation}, \bibinfo{year}{1994}.
\bibitem[{Tac et~al.(2022)Tac, Sree, Rausch, and Tepole}]{tac2022data}
\bibinfo{author}{V.~Tac}, \bibinfo{author}{V.~D. Sree}, \bibinfo{author}{M.~K.
  Rausch}, \bibinfo{author}{A.~B. Tepole},
\newblock \bibinfo{title}{Data-driven modeling of the mechanical behavior of
  anisotropic soft biological tissue},
\newblock \bibinfo{journal}{Engineering with Computers} \bibinfo{volume}{38}
  (\bibinfo{year}{2022}) \bibinfo{pages}{4167--4182}.
\bibitem[{Linka and Kuhl(2023)}]{linka2023new}
\bibinfo{author}{K.~Linka}, \bibinfo{author}{E.~Kuhl},
\newblock \bibinfo{title}{A new family of constitutive artificial neural
  networks towards automated model discovery},
\newblock \bibinfo{journal}{Computer Methods in Applied Mechanics and
  Engineering} \bibinfo{volume}{403} (\bibinfo{year}{2023})
  \bibinfo{pages}{115731}.
\bibitem[{Tac et~al.(2022)Tac, {Sahli Costabal}, and Tepole}]{TAC2022115248}
\bibinfo{author}{V.~Tac}, \bibinfo{author}{F.~{Sahli Costabal}},
  \bibinfo{author}{A.~B. Tepole},
\newblock \bibinfo{title}{Data-driven tissue mechanics with polyconvex neural
  ordinary differential equations},
\newblock \bibinfo{journal}{Computer Methods in Applied Mechanics and
  Engineering} \bibinfo{volume}{398} (\bibinfo{year}{2022})
  \bibinfo{pages}{115248}.
\bibitem[{Klein et~al.(2022)Klein, Fern{\'a}ndez, Martin, Neff, and
  Weeger}]{klein2022polyconvex}
\bibinfo{author}{D.~K. Klein}, \bibinfo{author}{M.~Fern{\'a}ndez},
  \bibinfo{author}{R.~J. Martin}, \bibinfo{author}{P.~Neff},
  \bibinfo{author}{O.~Weeger},
\newblock \bibinfo{title}{Polyconvex anisotropic hyperelasticity with neural
  networks},
\newblock \bibinfo{journal}{Journal of the Mechanics and Physics of Solids}
  \bibinfo{volume}{159} (\bibinfo{year}{2022}) \bibinfo{pages}{104703}.
\bibitem[{Amos et~al.(2017)Amos, Xu, and Kolter}]{amos2017input}
\bibinfo{author}{B.~Amos}, \bibinfo{author}{L.~Xu}, \bibinfo{author}{J.~Z.
  Kolter},
\newblock \bibinfo{title}{Input convex neural networks},
\newblock in: \bibinfo{booktitle}{International Conference on Machine
  Learning}, \bibinfo{organization}{PMLR}, pp. \bibinfo{pages}{146--155}.
\bibitem[{Ghaboussi and Sidarta(1998)}]{ghaboussi1998_2}
\bibinfo{author}{J.~Ghaboussi}, \bibinfo{author}{D.~Sidarta},
\newblock \bibinfo{title}{New nested adaptive neural networks (nann) for
  constitutive modeling},
\newblock \bibinfo{journal}{Computers and Geotechnics} \bibinfo{volume}{22}
  (\bibinfo{year}{1998}) \bibinfo{pages}{29--52}.
\bibitem[{Heider et~al.(2020)Heider, Wang, and Sun}]{heider2020so}
\bibinfo{author}{Y.~Heider}, \bibinfo{author}{K.~Wang},
  \bibinfo{author}{W.~Sun},
\newblock \bibinfo{title}{So (3)-invariance of informed-graph-based deep neural
  network for anisotropic elastoplastic materials},
\newblock \bibinfo{journal}{Computer Methods in Applied Mechanics and
  Engineering} \bibinfo{volume}{363} (\bibinfo{year}{2020})
  \bibinfo{pages}{112875}.
\bibitem[{Holzapfel(2000)}]{holzapfel2000}
\bibinfo{author}{G.~A. Holzapfel}, \bibinfo{title}{Nonlinear Solid Mechanics; A
  Continuum Approach for Engineering}, \bibinfo{publisher}{John Wiley \& Sons,
  LTD}, \bibinfo{year}{2000}.
\bibitem[{Ehret and Itskov(2007)}]{ehret2007polyconvex}
\bibinfo{author}{A.~E. Ehret}, \bibinfo{author}{M.~Itskov},
\newblock \bibinfo{title}{A polyconvex hyperelastic model for fiber-reinforced
  materials in application to soft tissues},
\newblock \bibinfo{journal}{Journal of Materials Science} \bibinfo{volume}{42}
  (\bibinfo{year}{2007}) \bibinfo{pages}{8853--8863}.
\bibitem[{Zhang and Garikipati(2020)}]{garikipati2020multiresolution}
\bibinfo{author}{X.~Zhang}, \bibinfo{author}{K.~Garikipati},
\newblock \bibinfo{title}{Machine learning materials physics: Multi-resolution
  neural networks learn the free energy and nonlinear elastic response of
  evolving microstructures},
\newblock \bibinfo{journal}{Computer Methods in Applied Mechanics and
  Engineering} \bibinfo{volume}{372} (\bibinfo{year}{2020})
  \bibinfo{pages}{113362}.
\bibitem[{Vlassis et~al.(2020)Vlassis, Ma, and Sun}]{vlassis2020geometric}
\bibinfo{author}{N.~N. Vlassis}, \bibinfo{author}{R.~Ma},
  \bibinfo{author}{W.~Sun},
\newblock \bibinfo{title}{Geometric deep learning for computational mechanics
  part i: Anisotropic hyperelasticity},
\newblock \bibinfo{journal}{Computer Methods in Applied Mechanics and
  Engineering} \bibinfo{volume}{371} (\bibinfo{year}{2020})
  \bibinfo{pages}{113299}.
\bibitem[{Liu et~al.(2020)Liu, Liang, and Sun}]{liu2020}
\bibinfo{author}{M.~Liu}, \bibinfo{author}{L.~Liang}, \bibinfo{author}{W.~Sun},
\newblock \bibinfo{title}{A generic physics-informed neural network-based
  constitutive model for soft biological tissues},
\newblock \bibinfo{journal}{Computer Methods in Applied Mechanics and
  Engineering} \bibinfo{volume}{372} (\bibinfo{year}{2020}).
\bibitem[{Fuhg et~al.(2022)Fuhg, Bouklas, and Jones}]{fuhg2022learning}
\bibinfo{author}{J.~N. Fuhg}, \bibinfo{author}{N.~Bouklas},
  \bibinfo{author}{R.~E. Jones},
\newblock \bibinfo{title}{Learning hyperelastic anisotropy from data via a
  tensor basis neural network},
\newblock \bibinfo{journal}{arXiv preprint arXiv:2204.04529}
  (\bibinfo{year}{2022}).
\bibitem[{Ball(1976)}]{ball1976convexity}
\bibinfo{author}{J.~M. Ball},
\newblock \bibinfo{title}{Convexity conditions and existence theorems in
  nonlinear elasticity},
\newblock \bibinfo{journal}{Archive for rational mechanics and Analysis}
  \bibinfo{volume}{63} (\bibinfo{year}{1976}) \bibinfo{pages}{337--403}.
\bibitem[{Schr{\"o}der(2010)}]{schroder2010anisotropic}
\bibinfo{author}{J.~Schr{\"o}der},
\newblock \bibinfo{title}{Anisotropic polyconvex energies},
\newblock in: \bibinfo{booktitle}{Poly-, quasi-and rank-one convexity in
  applied mechanics}, \bibinfo{publisher}{Springer}, \bibinfo{year}{2010}, pp.
  \bibinfo{pages}{53--105}.
\bibitem[{Schr{\"o}der and Neff(2003)}]{schroder2003invariant}
\bibinfo{author}{J.~Schr{\"o}der}, \bibinfo{author}{P.~Neff},
\newblock \bibinfo{title}{Invariant formulation of hyperelastic transverse
  isotropy based on polyconvex free energy functions},
\newblock \bibinfo{journal}{International journal of solids and structures}
  \bibinfo{volume}{40} (\bibinfo{year}{2003}) \bibinfo{pages}{401--445}.
\bibitem[{Gao et~al.(2017)Gao, Neff, Roventa, and Thiel}]{gao2017convexity}
\bibinfo{author}{D.~Y. Gao}, \bibinfo{author}{P.~Neff},
  \bibinfo{author}{I.~Roventa}, \bibinfo{author}{C.~Thiel},
\newblock \bibinfo{title}{On the convexity of nonlinear elastic energies in the
  right cauchy-green tensor},
\newblock \bibinfo{journal}{Journal of Elasticity} \bibinfo{volume}{127}
  (\bibinfo{year}{2017}) \bibinfo{pages}{303--308}.
\bibitem[{As'~ad et~al.(2022)As'~ad, Avery, and Farhat}]{as2022mechanics}
\bibinfo{author}{F.~As'~ad}, \bibinfo{author}{P.~Avery},
  \bibinfo{author}{C.~Farhat},
\newblock \bibinfo{title}{A mechanics-informed artificial neural network
  approach in data-driven constitutive modeling},
\newblock \bibinfo{journal}{International Journal for Numerical Methods in
  Engineering} \bibinfo{volume}{123} (\bibinfo{year}{2022})
  \bibinfo{pages}{2738--2759}.
\bibitem[{Chen and Guilleminot(2022)}]{CHEN2022103993}
\bibinfo{author}{P.~Chen}, \bibinfo{author}{J.~Guilleminot},
\newblock \bibinfo{title}{Polyconvex neural networks for hyperelastic
  constitutive models: A rectification approach},
\newblock \bibinfo{journal}{Mechanics Research Communications}
  \bibinfo{volume}{125} (\bibinfo{year}{2022}) \bibinfo{pages}{103993}.
\bibitem[{Lejeune(2020)}]{lejeune2020mechanical}
\bibinfo{author}{E.~Lejeune},
\newblock \bibinfo{title}{Mechanical mnist: A benchmark dataset for mechanical
  metamodels},
\newblock \bibinfo{journal}{Extreme Mechanics Letters} \bibinfo{volume}{36}
  (\bibinfo{year}{2020}) \bibinfo{pages}{100659}.
\bibitem[{Kobeissi et~al.(2022)Kobeissi, Mohammadzadeh, and
  Lejeune}]{kobeissi2022enhancing}
\bibinfo{author}{H.~Kobeissi}, \bibinfo{author}{S.~Mohammadzadeh},
  \bibinfo{author}{E.~Lejeune},
\newblock \bibinfo{title}{Enhancing mechanical metamodels with a generative
  model-based augmented training dataset},
\newblock \bibinfo{journal}{Journal of Biomechanical Engineering}
  \bibinfo{volume}{144} (\bibinfo{year}{2022}) \bibinfo{pages}{121002}.
\bibitem[{Dal et~al.(2021)Dal, A{\c{c}}{\i}kg{\"o}z, and
  Badienia}]{dal2021performance}
\bibinfo{author}{H.~Dal}, \bibinfo{author}{K.~A{\c{c}}{\i}kg{\"o}z},
  \bibinfo{author}{Y.~Badienia},
\newblock \bibinfo{title}{On the performance of isotropic hyperelastic
  constitutive models for rubber-like materials: a state of the art review},
\newblock \bibinfo{journal}{Applied Mechanics Reviews} \bibinfo{volume}{73}
  (\bibinfo{year}{2021}).
\bibitem[{Rus and Tolley(2015)}]{rus2015design}
\bibinfo{author}{D.~Rus}, \bibinfo{author}{M.~T. Tolley},
\newblock \bibinfo{title}{Design, fabrication and control of soft robots},
\newblock \bibinfo{journal}{Nature} \bibinfo{volume}{521}
  (\bibinfo{year}{2015}) \bibinfo{pages}{467--475}.
\bibitem[{Jor et~al.(2013)Jor, Parker, Taberner, Nash, and
  Nielsen}]{jor2013computational}
\bibinfo{author}{J.~W. Jor}, \bibinfo{author}{M.~D. Parker},
  \bibinfo{author}{A.~J. Taberner}, \bibinfo{author}{M.~P. Nash},
  \bibinfo{author}{P.~M. Nielsen},
\newblock \bibinfo{title}{Computational and experimental characterization of
  skin mechanics: identifying current challenges and future directions},
\newblock \bibinfo{journal}{Wiley Interdisciplinary Reviews: Systems Biology
  and Medicine} \bibinfo{volume}{5} (\bibinfo{year}{2013})
  \bibinfo{pages}{539--556}.
\bibitem[{Lanir and Fung(1974)}]{lanir1974two}
\bibinfo{author}{Y.~Lanir}, \bibinfo{author}{Y.~Fung},
\newblock \bibinfo{title}{Two-dimensional mechanical properties of rabbit
  skin—ii. experimental results},
\newblock \bibinfo{journal}{Journal of biomechanics} \bibinfo{volume}{7}
  (\bibinfo{year}{1974}) \bibinfo{pages}{171--182}.
\bibitem[{Lanir(1983)}]{lanir1983constitutive}
\bibinfo{author}{Y.~Lanir},
\newblock \bibinfo{title}{Constitutive equations for fibrous connective
  tissues},
\newblock \bibinfo{journal}{Journal of biomechanics} \bibinfo{volume}{16}
  (\bibinfo{year}{1983}) \bibinfo{pages}{1--12}.
\bibitem[{Toaquiza~Tubon et~al.(2022)Toaquiza~Tubon, Moreno-Flores, Sree, and
  Tepole}]{toaquiza2022anisotropic}
\bibinfo{author}{J.~D. Toaquiza~Tubon}, \bibinfo{author}{O.~Moreno-Flores},
  \bibinfo{author}{V.~D. Sree}, \bibinfo{author}{A.~B. Tepole},
\newblock \bibinfo{title}{Anisotropic damage model for collagenous tissues and
  its application to model fracture and needle insertion mechanics},
\newblock \bibinfo{journal}{Biomechanics and Modeling in Mechanobiology}
  \bibinfo{volume}{21} (\bibinfo{year}{2022}) \bibinfo{pages}{1--16}.
\bibitem[{Chen et~al.(2020)Chen, N{\'\i}~Annaidh, and
  Roccabianca}]{chen2020microstructurally}
\bibinfo{author}{S.~Chen}, \bibinfo{author}{A.~N{\'\i}~Annaidh},
  \bibinfo{author}{S.~Roccabianca},
\newblock \bibinfo{title}{A microstructurally inspired constitutive model for
  skin mechanics},
\newblock \bibinfo{journal}{Biomechanics and modeling in mechanobiology}
  \bibinfo{volume}{19} (\bibinfo{year}{2020}) \bibinfo{pages}{275--289}.
\bibitem[{Fuhg and Bouklas(2022)}]{fuhgPhysicsinformedDatadrivenIsotropic2022}
\bibinfo{author}{J.~N. Fuhg}, \bibinfo{author}{N.~Bouklas},
\newblock \bibinfo{title}{On physics-informed data-driven isotropic and
  anisotropic constitutive models through probabilistic machine learning and
  space-filling sampling},
\newblock \bibinfo{journal}{Computer Methods in Applied Mechanics and
  Engineering} \bibinfo{volume}{394} (\bibinfo{year}{2022})
  \bibinfo{pages}{114915}.
\bibitem[{Gasser et~al.(2005)Gasser, Ogden, and Holzapfel}]{gasser2005GOH}
\bibinfo{author}{T.~C. Gasser}, \bibinfo{author}{R.~W. Ogden},
  \bibinfo{author}{G.~A. Holzapfel},
\newblock \bibinfo{title}{Hyperelastic modelling of arterial layers with
  distributed collagen fibre orientations},
\newblock \bibinfo{journal}{Journal of the royal society interface}
  \bibinfo{volume}{3} (\bibinfo{year}{2005}) \bibinfo{pages}{15--35}.
\bibitem[{Fuhg et~al.(2022)Fuhg, Hamel, Johnson, Jones, and
  Bouklas}]{fuhgModularMachineLearningbased2022}
\bibinfo{author}{J.~N. Fuhg}, \bibinfo{author}{C.~M. Hamel},
  \bibinfo{author}{K.~Johnson}, \bibinfo{author}{R.~Jones},
  \bibinfo{author}{N.~Bouklas}, \bibinfo{title}{Modular machine learning-based
  elastoplasticity: Generalization in the context of limited data},
  \bibinfo{year}{2022}.
\bibitem[{Kirchdoerfer and Ortiz(2016)}]{kirchdoerfer2016data}
\bibinfo{author}{T.~Kirchdoerfer}, \bibinfo{author}{M.~Ortiz},
\newblock \bibinfo{title}{Data-driven computational mechanics},
\newblock \bibinfo{journal}{Computer Methods in Applied Mechanics and
  Engineering} \bibinfo{volume}{304} (\bibinfo{year}{2016})
  \bibinfo{pages}{81--101}.
\bibitem[{Ghaderi et~al.(2020)Ghaderi, Morovati, and
  Dargazany}]{ghaderiPhysicsInformedAssemblyFeedForward2020a}
\bibinfo{author}{A.~Ghaderi}, \bibinfo{author}{V.~Morovati},
  \bibinfo{author}{R.~Dargazany},
\newblock \bibinfo{title}{A {{Physics-Informed Assembly}} of {{Feed-Forward
  Neural Network Engines}} to {{Predict Inelasticity}} in {{Cross-Linked
  Polymers}}},
\newblock \bibinfo{journal}{Polymers} \bibinfo{volume}{12}
  (\bibinfo{year}{2020}) \bibinfo{pages}{2628}.
\bibitem[{Flaschel et~al.(2023)Flaschel, Kumar, and
  De~Lorenzis}]{flaschel2023automated}
\bibinfo{author}{M.~Flaschel}, \bibinfo{author}{S.~Kumar},
  \bibinfo{author}{L.~De~Lorenzis},
\newblock \bibinfo{title}{Automated discovery of generalized standard material
  models with euclid},
\newblock \bibinfo{journal}{Computer Methods in Applied Mechanics and
  Engineering} \bibinfo{volume}{405} (\bibinfo{year}{2023})
  \bibinfo{pages}{115867}.
\bibitem[{Tac et~al.(2021)Tac, Costabal, and Tepole}]{tac2021automatically}
\bibinfo{author}{V.~Tac}, \bibinfo{author}{F.~S. Costabal},
  \bibinfo{author}{A.~B. Tepole},
\newblock \bibinfo{title}{Automatically polyconvex strain energy functions
  using neural ordinary differential equations},
\newblock \bibinfo{journal}{arXiv preprint arXiv:2110.03774}
  (\bibinfo{year}{2021}).
\bibitem[{Vlassis and Sun(2021)}]{Vlassis202elastoplast}
\bibinfo{author}{N.~N. Vlassis}, \bibinfo{author}{W.~Sun},
\newblock \bibinfo{title}{Sobolev training of thermodynamic-informed neural
  networks for interpretable elasto-plasticity models with level set
  hardening},
\newblock \bibinfo{journal}{Computer Methods in Applied Mechanics and
  Engineering} \bibinfo{volume}{377} (\bibinfo{year}{2021}).
\bibitem[{Teichert et~al.(2019)Teichert, Natarajan, Van~der Ven, and
  Garikipati}]{teichert2019machine}
\bibinfo{author}{G.~H. Teichert}, \bibinfo{author}{A.~Natarajan},
  \bibinfo{author}{A.~Van~der Ven}, \bibinfo{author}{K.~Garikipati},
\newblock \bibinfo{title}{Machine learning materials physics: Integrable deep
  neural networks enable scale bridging by learning free energy functions},
\newblock \bibinfo{journal}{Computer Methods in Applied Mechanics and
  Engineering} \bibinfo{volume}{353} (\bibinfo{year}{2019})
  \bibinfo{pages}{201--216}.
\bibitem[{Steigmann(2003)}]{steigmann2003isotropic}
\bibinfo{author}{D.~J. Steigmann},
\newblock \bibinfo{title}{On isotropic, frame-invariant, polyconvex
  strain-energy functions},
\newblock \bibinfo{journal}{Quarterly Journal of Mechanics and Applied
  Mathematics} \bibinfo{volume}{56} (\bibinfo{year}{2003})
  \bibinfo{pages}{483--491}.
\bibitem[{Zhang et~al.(2022)Zhang, Sommer, Niestrawska, Holzapfel, and
  Nordsletten}]{zhang2022effects}
\bibinfo{author}{W.~Zhang}, \bibinfo{author}{G.~Sommer}, \bibinfo{author}{J.~A.
  Niestrawska}, \bibinfo{author}{G.~A. Holzapfel},
  \bibinfo{author}{D.~Nordsletten},
\newblock \bibinfo{title}{The effects of viscoelasticity on residual strain in
  aortic soft tissues},
\newblock \bibinfo{journal}{Acta Biomaterialia} \bibinfo{volume}{140}
  (\bibinfo{year}{2022}) \bibinfo{pages}{398--411}.
\bibitem[{Xu et~al.(2021)Xu, Tartakovsky, Burghardt, and
  Darve}]{xuLearningViscoelasticityModels2021}
\bibinfo{author}{K.~Xu}, \bibinfo{author}{A.~M. Tartakovsky},
  \bibinfo{author}{J.~Burghardt}, \bibinfo{author}{E.~Darve},
\newblock \bibinfo{title}{Learning viscoelasticity models from indirect data
  using deep neural networks},
\newblock \bibinfo{journal}{Computer Methods in Applied Mechanics and
  Engineering} \bibinfo{volume}{387} (\bibinfo{year}{2021})
  \bibinfo{pages}{114124}.
\bibitem[{Chen(2021)}]{chenRecurrentNeuralNetworks2021}
\bibinfo{author}{G.~Chen},
\newblock \bibinfo{title}{Recurrent neural networks ({{RNNs}}) learn the
  constitutive law of viscoelasticity},
\newblock \bibinfo{journal}{Computational Mechanics} \bibinfo{volume}{67}
  (\bibinfo{year}{2021}) \bibinfo{pages}{1009--1019}.
\bibitem[{Holzapfel et~al.(2021)Holzapfel, Linka, Sherifova, and
  Cyron}]{holzapfel2021predictive}
\bibinfo{author}{G.~A. Holzapfel}, \bibinfo{author}{K.~Linka},
  \bibinfo{author}{S.~Sherifova}, \bibinfo{author}{C.~J. Cyron},
\newblock \bibinfo{title}{Predictive constitutive modelling of arteries by deep
  learning},
\newblock \bibinfo{journal}{Journal of the Royal Society Interface}
  \bibinfo{volume}{18} (\bibinfo{year}{2021}) \bibinfo{pages}{20210411}.
\bibitem[{Leng et~al.(2021)Leng, Calve, and Tepole}]{leng2021}
\bibinfo{author}{Y.~Leng}, \bibinfo{author}{S.~Calve}, \bibinfo{author}{A.~B.
  Tepole},
\newblock \bibinfo{title}{Predicting the mechanical properties of fibrin using
  neural networks trained on discrete fiber network data},
\newblock \bibinfo{journal}{arXiv preprint arXiv:2101.11712}
  (\bibinfo{year}{2021}).
\bibitem[{Flaschel et~al.(2021)Flaschel, Kumar, and
  De~Lorenzis}]{flaschelUnsupervisedDiscoveryInterpretable2021}
\bibinfo{author}{M.~Flaschel}, \bibinfo{author}{S.~Kumar},
  \bibinfo{author}{L.~De~Lorenzis},
\newblock \bibinfo{title}{Unsupervised discovery of interpretable hyperelastic
  constitutive laws},
\newblock \bibinfo{journal}{Computer Methods in Applied Mechanics and
  Engineering} \bibinfo{volume}{381} (\bibinfo{year}{2021})
  \bibinfo{pages}{113852}.
\bibitem[{Chen and Gu(2021)}]{chen2021learning}
\bibinfo{author}{C.-T. Chen}, \bibinfo{author}{G.~X. Gu},
\newblock \bibinfo{title}{Learning hidden elasticity with deep neural
  networks},
\newblock \bibinfo{journal}{Proceedings of the National Academy of Sciences}
  \bibinfo{volume}{118} (\bibinfo{year}{2021}) \bibinfo{pages}{e2102721118}.

\end{thebibliography}
\end{document}